\documentclass[journal, hidelinks]{IEEEtran}

\usepackage{amsmath,amsfonts}
\usepackage{algorithmic}
\usepackage{algorithm}
\usepackage{array}
\usepackage[caption=false,font=normalsize,labelfont=sf,textfont=sf]{subfig}
\usepackage{textcomp}
\usepackage{stfloats}
\usepackage{url}
\usepackage{verbatim}
\usepackage{graphicx}
\usepackage{cite}
\usepackage{microtype}
\usepackage{amssymb,mathtools}
\usepackage{comment}
\usepackage{orcidlink}
\usepackage[switch]{lineno}
\usepackage{siunitx}
\usepackage{xfrac}
\usepackage{enumitem}

\usepackage[nameinlink,capitalize]{cleveref}
\crefname{figure}{Fig.}{Figs.}
\crefname{equation}{}{}
\crefname{section}{Sec.}{Secs.}

\usepackage{tikz}
\usepackage{tikz-3dplot}
\usepackage{circuitikz}
\usepackage{pgfplots}
\pgfplotsset{compat=1.16}
\pgfplotsset{compat=newest}
\usepgfplotslibrary{groupplots}
\usepgfplotslibrary{polar}
\usepgfplotslibrary{smithchart}
\usepgfplotslibrary{statistics}
\usepgfplotslibrary{dateplot}
\usepgfplotslibrary{ternary}
\usetikzlibrary{arrows.meta}
\usetikzlibrary{backgrounds}
\usepgfplotslibrary{patchplots}
\usepgfplotslibrary{fillbetween}
\usetikzlibrary{shapes.misc}
\tikzset{
	>=latex,
}

\usepackage[toc, acronym]{glossaries}
\glsdisablehyper

\newcommand{\ie}{\textit{i.e. }}

\newcommand{\cov}[2][]{\ensuremath{\mathbf{C}_{#2}^{\mathrm{#1}}}}
\newcommand{\norm}[2]{\ensuremath{\left\|#1\right\|_{\mathrm{#2}}}}
\newcommand{\herm}[1]{\ensuremath{#1^{\mathrm{H}}}}
\newcommand{\trans}[1]{\ensuremath{#1^{\mathrm{T}}}}
\newcommand{\expec}[2][]{\ensuremath{\mathrm{E}_{#1}\left[#2\right]}}
\newcommand{\trace}[1]{\ensuremath{\mathrm{tr}\left(#1\right)}}
\newcommand{\euler}{\ensuremath{\mathrm{e}}}
\newcommand{\im}{\ensuremath{\mathrm{j}}}

\newcommand{\id}[1]{\ensuremath{\mathbf{I}_{#1}}}
\newcommand{\bsf}[2][]{\ensuremath{\boldsymbol{\mathsf{#2}}^{\mathrm{#1}}}}
\newcommand{\cnumbers}{\ensuremath{\mathbb{C}}}
\newcommand{\rnumbers}{\ensuremath{\mathbb{R}}}
\newcommand{\znumbers}{\ensuremath{\mathbb{Z}}}

\newcommand{\dd}{\ensuremath{\mathop{\mathrm{d}\!}}}

\newcommand{\zeros}[1]{\ensuremath{\mathbf{0}_{#1}}}

\newcommand{\bolddot}{\ensuremath{\boldsymbol{\cdot}}}
\newcommand{\dist}[2][]{\ensuremath{d_{#2}^{\mathrm{#1}}}}

\newcommand{\sinr}[2]{\ensuremath{\mathrm{SINR}_{#1}^{#2}}}
\newcommand{\real}[1]{\ensuremath{\Re\left[#1\right]}}

\newcommand{\sigmamin}{\sigma_{\mathrm{min}}}

\newcommand{\R}[1]{\ensuremath{R_{\mathrm{#1}}}}
\newcommand{\X}[1]{\ensuremath{X_{\mathrm{#1}}}}
\newcommand{\Z}[1]{\ensuremath{Z_{\mathrm{#1}}}}

\newcommand{\bfZ}[2][]{\ensuremath{\mathbf{Z}_{\mathrm{#2}}^{\mathrm{#1}}}}
\newcommand{\bfDelta}[2][]{\ensuremath{\mathbf{\Delta}_{\mathrm{#2}}^{\mathrm{#1}}}}
\newcommand{\bsfZ}[2][]{\ensuremath{\boldsymbol{\mathsf{Z}}_{\mathrm{#2}}^{\mathrm{#1}}}}

\newcommand{\bsfvs}[1]{\ensuremath{\boldsymbol{\mathsf{v}}_{\mathrm{#1}}}}
\newcommand{\bsfis}[1]{\ensuremath{\boldsymbol{\mathsf{i}}_{\mathrm{#1}}}}
\newcommand{\bfis}[1]{\ensuremath{\mathbf{i}}_{\mathrm{#1}}}
\newcommand{\power}[1]{\ensuremath{P_{\mathrm{#1}}}}

\let\originalleft\left
\let\originalright\right
\renewcommand{\left}{\mathopen{}\mathclose\bgroup\originalleft}
\renewcommand{\right}{\aftergroup\egroup\originalright}

\newtheorem{definition}{Definition}
\newtheorem{proposition}{Proposition}
\newtheorem{lemma}{Lemma}
\newtheorem{corollary}{Corollary}

\begin{document}
    \newacronym{ap}{AP}{access point}
\newacronym{cms}{CMS}{canonical minimum scattering}
\newacronym{ehf}{EHF}{extremely high frequency}
\newacronym{lhs}{LHS}{left-hand side}
\newacronym{lna}{LNA}{low noise amplifier}
\newacronym{los}{LoS}{line-of-sight}
\newacronym{mimo}{MIMO}{multiple-input multiple-output}
\newacronym{mmse}{MMSE}{minimum mean square error}
\newacronym{mmtc}{mMTC}{massive machine-type communications}
\newacronym{nlos}{NLoS}{non-line-of-sight}
\newacronym{se}{SE}{spectral efficiency}
\newacronym{sinr}{SINR}{signal-to-interference-plus-noise ratio}
\newacronym{snr}{SNR}{signal-to-noise ratio}
\newacronym{thf}{THF}{tremendously high frequency}
\newacronym{ua}{UA}{unilateral approximation}
\newacronym{ula}{ULA}{uniform linear array}
\newacronym{upa}{UPA}{uniform planar array}
\newacronym{zf}{ZF}{zero-forcing}

    \bibliographystyle{IEEEtran-normspace}
    \bstctlcite{IEEEexample:BSTcontrol}

    \title{\huge Error Bounds and Necessary Conditions for the Unilateral Approximation and Their Impact on Spectral Efficiency}

    \author{Aniol Martí\,\orcidlink{0000-0002-5600-8541},~\IEEEmembership{Member,~IEEE},
		Antonio A. D'Amico\,\orcidlink{0000-0002-4017-7439},~\IEEEmembership{Senior~Member,~IEEE},\\
		Luca Sanguinetti\,\orcidlink{0000-0002-2577-4091},~\IEEEmembership{Fellow,~IEEE},
		and Xavier Gràcia\,\orcidlink{0000-0003-1006-4086}
		\thanks{This work was supported by grant 2025 SGR 00902 from the Departament de Recerca i Universitats de la Generalitat de Catalunya, and by the Italian Ministry of Education and Research (MUR) within the framework of the FoReLab Project (Department of Excellence).}%
		\thanks{A. Martí is with the Department of Signal Theory and Communications, Universitat Politècnica de Catalunya, Barcelona, Catalonia, Spain (e-mail: aniol.marti@upc.edu).}
		\thanks{A. A. D'Amico and L. Sanguinetti are with the Dipartimento di Ingegneria dell'Informazione, University of Pisa, Pisa, Italy (e-mail: antonio.damico@unipi.it, luca.sanguinetti@unipi.it).}
		\thanks{X. Gràcia is with the Department of Mathematics, Universitat Politècnica de Catalunya, Barcelona, Catalonia, Spain (e-mail: xavier.gracia@upc.edu).}
	}

    \maketitle

    \begin{abstract}
        This paper investigates the validity of the unilateral approximation in multiuser near-field \acrlong{mimo} wireless communication systems employing large \acrlongpl{upa}.
        We first establish that circuit-theoretic consistency requires satisfying both input-side and output-side conditions simultaneously.
        Under perturbation-based sufficient conditions, we derive closed-form upper bounds on the relative errors incurred in the physical channel, power coupling, and noise covariance matrices.
        Furthermore, combining electromagnetic Hertzian dipole formulations with continuous 2D spatial Fourier analysis, we mathematically prove that scaling the antenna array aperture and element count does not inherently violate the unilateral assumption in the asymptotic regime.
        From a communication perspective, we show that mismatched linear receive combiners (such as zero-forcing) experience severe near-field spectral efficiency degradation in multiuser environments employing either Hertzian or half-wavelength dipoles.
        Consequently, circuit-level equivalence does not guarantee signal processing robustness, highlighting the need for coupling-aware receiver processing and channel modeling in near-field communication systems.
    \end{abstract}


    \section{Introduction}
        \IEEEPARstart{W}{ireless} communications have become ubiquitous in today’s increasingly connected world, spanning a wide range of propagation distances, from very short-range scenarios such as on-chip communications, through intermediate-range applications including wireless data centers, \acrfull{mmtc}, and mobile communications, to very long-range links such as satellite communications~\cite{rappaport_wireless_2009,abadal_wave_2020,hamza_wireless_2016}.

        These diverse applications require wireless systems not only to provide increasingly high data rates, but also to support a massive number of heterogeneous devices while meeting stringent latency and reliability requirements.
        To address these challenges, two complementary strategies have emerged: the use of extremely large (and potentially dense) antenna arrays and operation in the \acrfull{ehf} and \acrfull{thf} bands~\cite{chafii_twelve_2023}.
        However, these technologies also introduce significant challenges in system design, implementation, and analysis.

        On the one hand, very short-range applications, as well as increasing array apertures and operating frequencies, make near-field operation increasingly relevant.
        Indeed, increasing the array aperture or operating frequency extends the far-field boundary (typically, the Fraunhofer distance) to several hundreds or even thousands of meters.
        Consequently, newly proposed communication schemes must account for the distinctive characteristics of the near field, most notably the spherical wavefront~\cite{lu_tutorial_2024}.
        On the other hand, short distances between radiating elements, such as those between the transmitter and receiver in short-range communications or between antenna elements in holographic \acrfull{mimo}, result in significant mutual coupling~\cite[Sec.~8.7]{balanis_antenna_2016}.
        In the literature, particularly in the context of mobile communications, only intra-array mutual coupling is typically taken into account, whereas the interaction between the transmitter and receiver is modeled exclusively as the intended propagation channel, neglecting possible reflected components.

        Several approaches have been proposed to address the challenges imposed by the aforementioned phenomena~\cite{mezghani_reincorporating_2024,zhu_electromagnetic_2024}.
        Among them, \textit{multiport communication theory} has emerged as a particularly promising framework~\cite{ivrlac_toward_2010}.
        This approach adopts a circuit-theoretic perspective in which the multiple-antenna communication system is modeled as a multiport black box characterized by its impedance or scattering matrix.
        The multiport framework is being widely employed in literature to model mutually coupled systems and characterize different aspects such as array gain~\cite{laas_limits_2020}, uplink/downlink reciprocity~\cite{damico_holographic_2024,laas_reciprocity_2020}, and the impact of mutual coupling on the \acrlong{se}~(\acrshort{se})~\cite{lu_compact_2026}.
        Most of these works, however, assume that there is no coupling from receiver to transmitter, something referred to as \textit{\acrlong{ua}}.
        Although this assumption is well justified in far-field communications~\cite{ivrlac_toward_2010}, its validity is far less evident in the near field, particularly in short-range scenarios.
        Notable exceptions include~\cite{laas_limits_2020}, where the authors analyze the limits of array gain while accounting for receiver-to-transmitter coupling.

        Motivated by this observation, some works have investigated the validity of the \acrlong{ua}.
        In~\cite{marti_asymptotic_2024,marti_unilateral_2026}, the authors consider a single user communicating with a \acrfull{ula} and assess the validity of the \acrlong{ua} from a circuit-theoretic perspective under \acrfull{los} and \acrfull{nlos} propagation, respectively.
        Similarly, in~\cite{qian_physically_2026}, a \acrshort{ula} of half-wavelength dipoles communicating with a single user is considered.
        The achievable rate of a system operating under the \acrlong{ua} is compared with that of an equivalent system accounting for receiver-to-transmitter coupling.

        The brief state-of-the-art review above shows that, to the best of the authors' knowledge, no previous work has investigated the \acrlong{ua} in planar arrays or multiuser settings, nor has compared the \acrshort{se} of the fully coupled system with that of a mismatched system.
        This work addresses these open problems in large arrays and near-field \acrshort{los} communications by first discussing the foundations of the \acrlong{ua}, then conducting an analytical and numerical investigation for arrays of Hertzian dipoles, and finally verifying the results through practical antenna models such as half-wavelength dipoles.

        We now provide an outline of the paper structure, highlighting its main contributions:
        \begin{itemize}
            \item In \cref{sec:multiport}, we review the multiport communication theory framework, with special emphasis on the unilateral approximation and its importance in system design and implementation.

            \item In \cref{sec:problem_statement}, we discuss the definition of the unilateral approximation, arguing that not only the state-of-the-art condition but also an analogous output-side condition must be fulfilled.
            The considered multiuser system is also defined in this section.

            \item In \cref{sec:error_bounds}, we derive analytical bounds for the error incurred by the unilateral approximation in the main matrices of the multiport model: namely, the channel, power coupling, and noise covariance matrices.

            \item In \cref{sec:analysis_hertzian_dipoles}, we combine Hertzian dipole models with spatial Fourier analysis to investigate the validity of the unilateral approximation in large arrays from a circuit-theoretic perspective.

            \item In \cref{sec:numerical_analysis}, the theoretical results obtained in the previous two sections are verified numerically in practical communication scenarios.

            \item In \cref{sec:spectral_efficiency}, we simulate the impact of the unilateral approximation on the \acrshort{se} of multiuser systems employing either Hertzian or half-wavelength dipoles and compare it with the circuit-theoretical results.
        \end{itemize}

        The following notation is used throughout the text.
        Boldface lowercase and uppercase letters denote vectors and matrices.
        $\Re[\cdot]$ extracts the real part of a complex-valued argument.
        Given a matrix $\mathbf{A}$, the element in its row $r$ and column $c$ is denoted as $[\mathbf{A}]_{r,c}$.
        The transpose operator is $\trans{\cdot}$ and the conjugate transpose one is $\herm{\cdot}$.
        The trace of a matrix $\mathbf{A}$ is $\trace{\mathbf{A}}$, while its Frobenius and spectral norms are denoted by $\norm{\mathbf{A}}{F}$ and $\norm{\mathbf{A}}{2}$, respectively.
        The minimum singular value and minimum eigenvalue of a matrix are denoted by $\sigmamin(\cdot)$ and $\lambda_{\mathrm{min}}(\cdot )$.
        The set of $MN \times MN$ block-Toeplitz matrices with $N \times N$ Toeplitz blocks is denoted by $\mathcal{BTTB}_{M, N}$.
        The $N$-dimensional unitary Fourier matrix is denoted by $\mathbf{F}_N$, whereas the continuous 2D Fourier transform is $\mathcal{F}_{\mathrm{2D}}[\cdot]$.
        Expectation is denoted by $\expec[]{\cdot}$.
        The impedance of free space is denoted by $\eta$, and the angular wavenumber is defined as $\kappa=2\pi/\lambda$, with $\lambda$ the wavelength.

    \section{Multiport Communication Theory}
        \label{sec:multiport}

        We now review the multiport communication model.
        Consider a narrowband communication system equipped with $K$ transmit antennas and $N$ receive antennas.
        To normalize the physical quantities into information-theoretic symbols $\bsf{x} \in \mathbb{C}^K$, we define the average transmitted power as $\power{T} = \expec{\bsf[H]{x}\bsf{x}}$.
        The discrete baseband representation of the received signal is given by
        \begin{equation}
            \label{eq:mimo}
            \bsf{y} = \bsf{H}\bsf{x} + \bsf{z},\quad \bsf{x}\in\cnumbers^{K},\ \bsf{y}\in\cnumbers^{N},
        \end{equation}
        where $\bsf{z}\in\cnumbers^{N}$ is additive Gaussian noise and $\bsf{H}\in\cnumbers^{N \times K}$ is the \acrshort{mimo} channel matrix encapsulating the physical model.
        Multiport communication theory provides a circuit-theoretic approach for deriving physically consistent expressions for all quantities in \cref{eq:mimo}.

\begin{figure}[ht!]
            \centering
            \resizebox{0.9\columnwidth}{!}{%
                \begin{tikzpicture}[scale=2.54]%
\ifx\dpiclw\undefined\newdimen\dpiclw\fi
\global\def\dpicdraw{\draw[line width=\dpiclw]}
\global\def\dpicstop{;}
\dpiclw=0.8bp
\dpiclw=0.8bp
\dpicdraw (0.395,-0.984252) rectangle (1.123346,0.984252)\dpicstop
\draw (0.759173,0) node{$\boldsymbol{\mathsf{Z}}_{\mathrm{A}}$};
\dpicdraw (0.395,0.752663)
 --(0.02,0.752663)\dpicstop
\dpicdraw[fill=white](0.02,0.752663) circle (0.007874in)\dpicstop
\dpicdraw (0.395,0.40528)
 --(0.02,0.40528)\dpicstop
\dpicdraw[fill=white](0.02,0.40528) circle (0.007874in)\dpicstop
\dpicdraw (0.395,-0.40528)
 --(0.02,-0.40528)\dpicstop
\dpicdraw[fill=white](0.02,-0.40528) circle (0.007874in)\dpicstop
\dpicdraw (0.395,-0.752663)
 --(0.02,-0.752663)\dpicstop
\dpicdraw[fill=white](0.02,-0.752663) circle (0.007874in)\dpicstop
\dpicdraw (1.123346,0.752663)
 --(1.498346,0.752663)\dpicstop
\dpicdraw[fill=white](1.498346,0.752663) circle (0.007874in)\dpicstop
\dpicdraw (1.123346,0.40528)
 --(1.498346,0.40528)\dpicstop
\dpicdraw[fill=white](1.498346,0.40528) circle (0.007874in)\dpicstop
\dpicdraw (1.123346,-0.40528)
 --(1.498346,-0.40528)\dpicstop
\dpicdraw[fill=white](1.498346,-0.40528) circle (0.007874in)\dpicstop
\dpicdraw (1.123346,-0.752663)
 --(1.498346,-0.752663)\dpicstop
\dpicdraw[fill=white](1.498346,-0.752663) circle (0.007874in)\dpicstop
\dpicdraw (-0,0.752663)
 --(-0.19375,0.752663)\dpicstop
\dpicdraw (-0.44375,0.752663)
 --(-0.44375,0.702663)
 --(-0.19375,0.702663)
 --(-0.19375,0.802663)
 --(-0.44375,0.802663)
 --(-0.44375,0.752663)\dpicstop
\dpicdraw (-0.44375,0.752663)
 --(-0.6375,0.752663)\dpicstop
\draw (-0.31875,0.802663) node[above=-2bp]{$ Z_{\mathrm{G}}$};
\filldraw[line width=0bp](0.134375,0.727663)
 --(0.234375,0.752663)
 --(0.134375,0.777663) --cycle\dpicstop
\dpicdraw (0.211469,0.752663)
 --(0.134375,0.752663)\dpicstop
\draw (0.172922,0.752663) node[above=-2bp]{$ \mathsf{i}_{\mathrm{T},1}$};
\dpicdraw (-0.631944,0.752663)
 --(-0.643056,0.752663)\dpicstop
\dpicdraw (-0.6375,0.40528)
 --(-0.6375,0.453972)\dpicstop
\dpicdraw (-0.6375,0.578972) circle (0.049213in)\dpicstop
\draw (-0.6375,0.516472) node{\scriptsize$-$};
\draw (-0.6375,0.641472) node{\scriptsize$+$};
\dpicdraw (-0.6375,0.703972)
 --(-0.6375,0.752663)\dpicstop
\draw (-0.7625,0.578972) node[left=-2bp]{$ \mathsf{v}_{\mathrm{G},1}$};
\dpicdraw (-0.6375,0.410836)
 --(-0.6375,0.399725)\dpicstop
\dpicdraw (-0.6375,0.40528)
 --(-0,0.40528)\dpicstop
\dpicdraw (-0,-0.40528)
 --(-0.19375,-0.40528)\dpicstop
\dpicdraw (-0.44375,-0.40528)
 --(-0.44375,-0.45528)
 --(-0.19375,-0.45528)
 --(-0.19375,-0.35528)
 --(-0.44375,-0.35528)
 --(-0.44375,-0.40528)\dpicstop
\dpicdraw (-0.44375,-0.40528)
 --(-0.6375,-0.40528)\dpicstop
\draw (-0.31875,-0.35528) node[above=-2bp]{$ Z_{\mathrm{G}}$};
\filldraw[line width=0bp](0.134375,-0.43028)
 --(0.234375,-0.40528)
 --(0.134375,-0.38028) --cycle\dpicstop
\dpicdraw (0.211469,-0.40528)
 --(0.134375,-0.40528)\dpicstop
\draw (0.172922,-0.40528) node[above=-2bp]{$ \mathsf{i}_{\mathrm{T},K}$};
\dpicdraw (-0.631944,-0.40528)
 --(-0.643056,-0.40528)\dpicstop
\dpicdraw (-0.6375,-0.752663)
 --(-0.6375,-0.703972)\dpicstop
\dpicdraw (-0.6375,-0.578972) circle (0.049213in)\dpicstop
\draw (-0.6375,-0.641472) node{\scriptsize$-$};
\draw (-0.6375,-0.516472) node{\scriptsize$+$};
\dpicdraw (-0.6375,-0.453972)
 --(-0.6375,-0.40528)\dpicstop
\draw (-0.7625,-0.578972) node[left=-2bp]{$ \mathsf{v}_{\mathrm{G},K}$};
\dpicdraw (-0.6375,-0.747108)
 --(-0.6375,-0.758219)\dpicstop
\dpicdraw (-0.6375,-0.752663)
 --(-0,-0.752663)\dpicstop
\dpicdraw[fill=white](0.02,0.752663) circle (0.007874in)\dpicstop
\dpicdraw[fill=white](0.02,0.40528) circle (0.007874in)\dpicstop
\draw (0.02,0.683972) node{$ +$};
\draw (0.02,0.578972) node{$ \mathsf{v}_{\mathrm{T},1}$};
\draw (0.02,0.473972) node{$ -$};
\dpicdraw[fill=white](0.02,-0.40528) circle (0.007874in)\dpicstop
\dpicdraw[fill=white](0.02,-0.752663) circle (0.007874in)\dpicstop
\draw (0.02,-0.473972) node{$ +$};
\draw (0.02,-0.578972) node{$ \mathsf{v}_{\mathrm{T},K}$};
\draw (0.02,-0.683972) node{$ -$};
\dpicdraw (2.018346,0.752663)
 --(1.893346,0.752663)\dpicstop
\dpicdraw (1.768346,0.752663) circle (0.049213in)\dpicstop
\draw (1.830846,0.752663) node{\scriptsize$-$};
\draw (1.705846,0.752663) node{\scriptsize$+$};
\dpicdraw (1.643346,0.752663)
 --(1.518346,0.752663)\dpicstop
\draw (1.768346,0.877663) node[above=-2bp]{$ \mathsf{v}_{\mathrm{N},1}$};
\filldraw[line width=0bp](1.380846,0.777663)
 --(1.280846,0.752663)
 --(1.380846,0.727663) --cycle\dpicstop
\dpicdraw (1.303753,0.752663)
 --(1.380846,0.752663)\dpicstop
\draw (1.3423,0.752663) node[above=-2bp]{$ \mathsf{i}_{\mathrm{R},1}$};
\dpicdraw (2.018346,0.752663)
 --(2.093346,0.752663)\dpicstop
\dpicdraw (2.093346,0.40528)
 --(2.093346,0.453972)\dpicstop
\dpicdraw (2.093346,0.578972) circle (0.049213in)\dpicstop
\filldraw[line width=0bp](2.118346,0.572722)
 --(2.093346,0.672722)
 --(2.068346,0.572722) --cycle\dpicstop
\dpicdraw (2.093346,0.485222)
 --(2.093346,0.649816)\dpicstop
\dpicdraw (2.093346,0.703972)
 --(2.093346,0.752663)\dpicstop
\draw (2.093346,0.29278) node{$\mathsf{i}_{\mathrm{LNA},1}$};
\dpicdraw (2.093346,0.40528)
 --(1.518346,0.40528)\dpicstop
\dpicdraw (2.093346,0.752663)
 --(2.543346,0.752663)\dpicstop
\dpicdraw (2.537791,0.752663)
 --(2.548902,0.752663)\dpicstop
\dpicdraw (2.543346,0.752663)
 --(2.543346,0.703972)\dpicstop
\dpicdraw (2.543346,0.453972)
 --(2.593346,0.453972)
 --(2.593346,0.703972)
 --(2.493346,0.703972)
 --(2.493346,0.453972)
 --(2.543346,0.453972)\dpicstop
\dpicdraw (2.543346,0.453972)
 --(2.543346,0.40528)\dpicstop
\draw (2.493346,0.578972) node[left=-2bp]{$ Z_{\mathrm{L}}$};
\draw (2.805846,0.728972) node{$ +$};
\draw (2.805846,0.578972) node{$ \mathsf{v}_{\mathrm{L},1}$};
\draw (2.805846,0.428972) node{$ -$};
\dpicdraw (2.543346,0.410836)
 --(2.543346,0.399725)\dpicstop
\dpicdraw (2.543346,0.40528)
 --(2.093346,0.40528)\dpicstop
\dpicdraw[fill=white](1.498346,0.752663) circle (0.007874in)\dpicstop
\dpicdraw[fill=white](1.498346,0.40528) circle (0.007874in)\dpicstop
\draw (1.498346,0.683972) node{$ +$};
\draw (1.498346,0.578972) node{$ \mathsf{v}_{\mathrm{R},1}$};
\draw (1.498346,0.473972) node{$ -$};
\dpicdraw (2.018346,-0.40528)
 --(1.893346,-0.40528)\dpicstop
\dpicdraw (1.768346,-0.40528) circle (0.049213in)\dpicstop
\draw (1.830846,-0.40528) node{\scriptsize$-$};
\draw (1.705846,-0.40528) node{\scriptsize$+$};
\dpicdraw (1.643346,-0.40528)
 --(1.518346,-0.40528)\dpicstop
\draw (1.768346,-0.28028) node[above=-2bp]{$ \mathsf{v}_{\mathrm{N},N}$};
\filldraw[line width=0bp](1.380846,-0.38028)
 --(1.280846,-0.40528)
 --(1.380846,-0.43028) --cycle\dpicstop
\dpicdraw (1.303753,-0.40528)
 --(1.380846,-0.40528)\dpicstop
\draw (1.3423,-0.40528) node[above=-2bp]{$ \mathsf{i}_{\mathrm{R},N}$};
\dpicdraw (2.018346,-0.40528)
 --(2.093346,-0.40528)\dpicstop
\dpicdraw (2.093346,-0.752663)
 --(2.093346,-0.703972)\dpicstop
\dpicdraw (2.093346,-0.578972) circle (0.049213in)\dpicstop
\filldraw[line width=0bp](2.118346,-0.585222)
 --(2.093346,-0.485222)
 --(2.068346,-0.585222) --cycle\dpicstop
\dpicdraw (2.093346,-0.672722)
 --(2.093346,-0.508128)\dpicstop
\dpicdraw (2.093346,-0.453972)
 --(2.093346,-0.40528)\dpicstop
\draw (2.093346,-0.865163) node{$\mathsf{i}_{\mathrm{LNA},N}$};
\dpicdraw (2.093346,-0.752663)
 --(1.518346,-0.752663)\dpicstop
\dpicdraw (2.093346,-0.40528)
 --(2.543346,-0.40528)\dpicstop
\dpicdraw (2.537791,-0.40528)
 --(2.548902,-0.40528)\dpicstop
\dpicdraw (2.543346,-0.40528)
 --(2.543346,-0.453972)\dpicstop
\dpicdraw (2.543346,-0.703972)
 --(2.593346,-0.703972)
 --(2.593346,-0.453972)
 --(2.493346,-0.453972)
 --(2.493346,-0.703972)
 --(2.543346,-0.703972)\dpicstop
\dpicdraw (2.543346,-0.703972)
 --(2.543346,-0.752663)\dpicstop
\draw (2.493346,-0.578972) node[left=-2bp]{$ Z_{\mathrm{L}}$};
\draw (2.805846,-0.428972) node{$ +$};
\draw (2.805846,-0.578972) node{$ \mathsf{v}_{\mathrm{L},N}$};
\draw (2.805846,-0.728972) node{$ -$};
\dpicdraw (2.543346,-0.747108)
 --(2.543346,-0.758219)\dpicstop
\dpicdraw (2.543346,-0.752663)
 --(2.093346,-0.752663)\dpicstop
\dpicdraw[fill=white](1.498346,-0.40528) circle (0.007874in)\dpicstop
\dpicdraw[fill=white](1.498346,-0.752663) circle (0.007874in)\dpicstop
\draw (1.498346,-0.473972) node{$ +$};
\draw (1.498346,-0.578972) node{$ \mathsf{v}_{\mathrm{R},N}$};
\draw (1.498346,-0.683972) node{$ -$};
\dpicdraw[fill=black](0.02,0.059055) circle (0.007874in)\dpicstop
\dpicdraw[fill=black](0.02,0.171555) circle (0.007874in)\dpicstop
\dpicdraw[fill=black](0.02,-0.053445) circle (0.007874in)\dpicstop
\dpicdraw[fill=black](1.498346,0.059055) circle (0.007874in)\dpicstop
\dpicdraw[fill=black](1.498346,0.171555) circle (0.007874in)\dpicstop
\dpicdraw[fill=black](1.498346,-0.053445) circle (0.007874in)\dpicstop
\filldraw[line width=0bp](0.125,-0.935144)
 --(0.225,-0.910144)
 --(0.125,-0.885144) --cycle\dpicstop
\dpicdraw (0.202094,-0.910144)
 --(-0.089961,-0.910144)
 --(-0.089961,-1.285144)\dpicstop
\draw (-0.089961,-1.285144) node[below=-2bp]{$\boldsymbol{\mathsf{Z}}_\mathrm{in}$};
\filldraw[line width=0bp](1.393346,-0.885144)
 --(1.293346,-0.910144)
 --(1.393346,-0.935144) --cycle\dpicstop
\dpicdraw (1.316253,-0.910144)
 --(1.608307,-0.910144)
 --(1.608307,-1.285144)\dpicstop
\draw (1.608307,-1.285144) node[below=-2bp]{$\boldsymbol{\mathsf{Z}}_\mathrm{out}$};
\end{tikzpicture}%
            }
            \caption{Circuit-theoretic multiport model of a $K \times N$ multi-antenna system.}
            \label{fig:multiport_mimo}
        \end{figure}
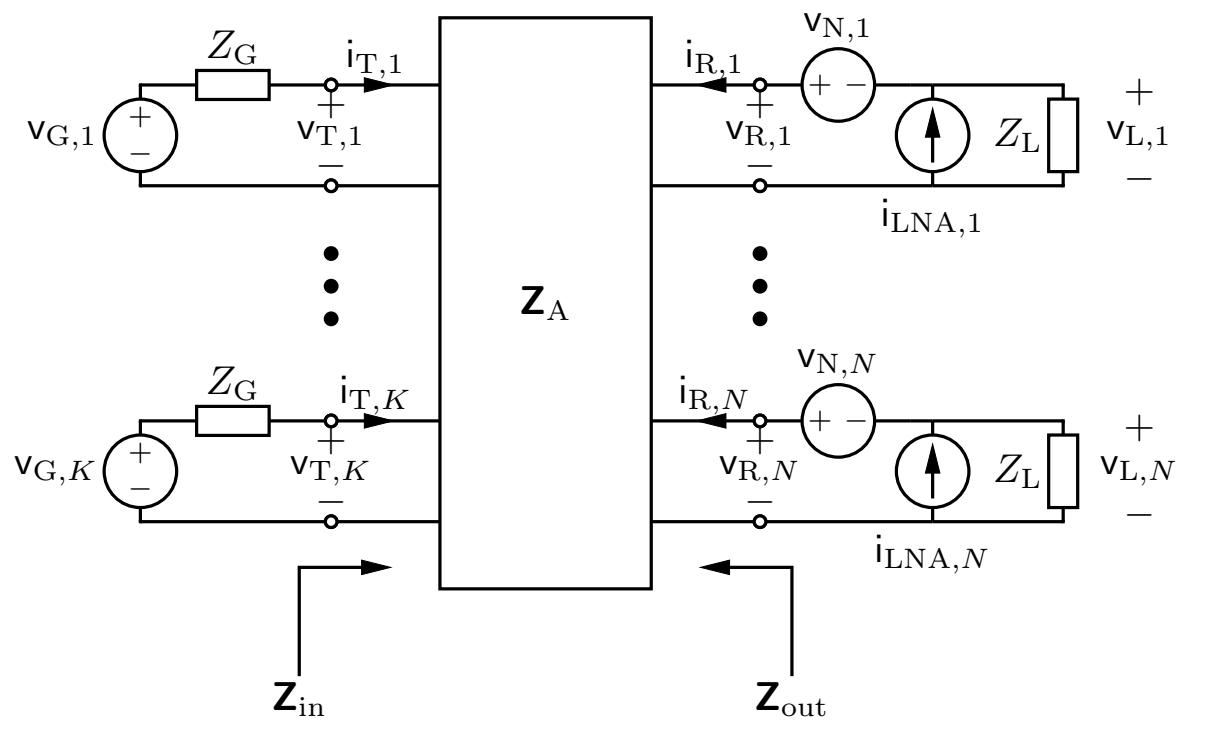

        The multiport model is depicted in~\cref{fig:multiport_mimo}.
        Although impedance matching networks are usually included, they are omitted here for two reasons: first, no closed-form expressions exist in the presence of receiver-to-transmitter coupling~\cite{phang_near-field_2018,mezghani_reincorporating_2024}; second, they become impractical to implement in large-scale arrays~\cite{laas_limits_2020,laas_reciprocity_2020}.

        Throughout this paper, we employ the circuit variables defined in~\cref{fig:multiport_mimo}, together with their vectorized representations. For example,
        \begin{equation}
			\bsfvs{T} = \trans{\left(\mathsf{v}_{\mathrm{T},1}, \dots, \mathsf{v}_{\mathrm{T},K}\right)},\quad \bsfis{T} = \trans{\left(\mathsf{i}_{\mathrm{T},1}, \dots, \mathsf{i}_{\mathrm{T},K}\right)}.
		\end{equation}
        Unless otherwise stated, all voltages and currents denote complex envelopes~\cite[Ch.~2]{madhow_introduction_2014}.
        Equivalent impedances are denoted by an arrow pointing from the computation point toward the corresponding port.

        \subsection{Antenna Mutual Coupling}
            The antenna multiport $\bsfZ{A} \in \cnumbers^{(K+N)\times(K+N)}$ is a matrix, generally random, that models mutual coupling and can be partitioned as
            \begin{equation}
				\label{eq:z_a}
				\begin{pmatrix}
					\bsfvs{T}\\
					\bsfvs{R}
				\end{pmatrix} =
				\begin{pmatrix}
					\bfZ{T} & \bsfZ{TR}\\
					\bsfZ{RT} & \bfZ{R}
				\end{pmatrix}
				\begin{pmatrix}
					\bsfis{T}\\
					\bsfis{R}
				\end{pmatrix},
			\end{equation}
            where $\bfZ{T} \in \cnumbers^{K \times K}$ and $\bfZ{R} \in \cnumbers^{N \times N}$ describe intra-array coupling at the transmit and receive sides, respectively, while $\bsfZ{TR} \in \cnumbers^{K \times N}$ and $\bsfZ{RT} \in \cnumbers^{N \times K}$ model inter-array coupling between the transmit and receive arrays.
            Taking into account antenna reciprocity~\cite[Sec.~3.8]{balanis_antenna_2016}, it follows that $\bfZ{T} = \bfZ[T]{T}$, $\bfZ{R} = \bfZ[T]{R}$, and $\bsfZ{TR} = \bsfZ[T]{RT}$.

            The equivalent input and output impedances of the antenna coupling network are given by~\cite{marti_physically_2026}:
                \begin{equation}
                    \label{eq:zin_zout}
                    \begin{aligned}
                        &\bsfZ{in} = \bfZ{T}-\bsfZ{TR}\left(\Z{L}\id{N}+\bfZ{R}\right)^{-1}\bsfZ{RT} = \bfZ{T}-\bfDelta[]{in},\\
                        &\bsfZ{out} = \bfZ{R}-\bsfZ{RT}\left(\Z{G}\id{K}+\bfZ{T}\right)^{-1}\bsfZ{TR} = \bfZ{R}-\bfDelta[]{out}.
                    \end{aligned}
                \end{equation}

        \subsection{Noise}
            The multiport communication framework considers both extrinsic and intrinsic noise sources.
            The former, originated by background radiation, is modeled by voltage sources connected to the receiver antenna ports when no current flows (\ie open-circuit)~\cite{ivrlac_toward_2010, damico_holographic_2024}.
            The intrinsic noise is generated by the subsystems following the receive matching network.
            Since it mainly originates in the \acrfullpl{lna}~\cite{ivrlac_toward_2010,damico_holographic_2024}, it is modeled by $N$ voltage and $N$ current sources.

            All sources are assumed Gaussian zero-mean random variables with the following statistics: $\mathrm{E}_{}[\bsfis{LNA}\herm{\bsfis{LNA}}] = \sigma_i^2\id{N}$, $\mathrm{E}_{}[\bsfvs{N}\herm{\bsfis{LNA}}] = \rho\R{N}\sigma_i^2 \id{N}$ and
            \begin{equation}
                \mathrm{E}_{}[\bsfvs{N}\herm{\bsfvs{N}}] = \underbrace{\R{N}^2\sigma_i^2\id{N}}_{\text{Intrinsic noise}} + \underbrace{4k_{\mathrm{B}}T_{\mathrm{A}}B_{\mathrm{W}} \real{\bsfZ{out}}}_{\text{Extrinsic noise}}.
            \end{equation}
            In the previous expressions, $k_{\mathrm{B}}$ is the Boltzmann constant, $T_{\mathrm{A}}$ is the noise temperature of the antennas, $B_{\mathrm{W}}$ is the equivalent noise bandwidth, $\R{N}$ and $\rho$ denote the noise resistance and the complex noise correlation coefficient of the \acrshortpl{lna}, and $\sigma_i^2 = 4 k_{\mathrm{B}} B_{\mathrm{W}} T_{\mathrm{A}}/\R{N}$.

        \subsection{Transmit power}
            \label{sec:power_noise_matrices}

            Transmit power is defined as the average active power delivered to the transmitting antennas:
            \begin{equation}
				\label{eq:physical_power}
				\power{T} = \expec{\real{\herm{\bsfvs{T}}\bsfis{T}}}.
			\end{equation}

            From Ohm’s law and the multiport relations we have $\bsfvs{T} = \bsfZ{in}\bsfis{T}$ and $\bsfis{T} = \left(\Z{G}\id{K}+\bsfZ{in}\right)^{-1}\bsfvs{G}$.
            Hence, the transmit power can be expressed in terms of the signal generators as
            \begin{equation}
                \label{eq:power_T}
                \power{T} = \frac{\expec{\herm{\bsfvs{G}}\bsf{B}\bsfvs{G}}{}}{\R{G}}
            \end{equation}
            where $\R{G} = \real{\Z{G}}$ and $\bsf{B}$ denotes the power coupling matrix given by
            \begin{equation}
                \label{eq:power_coupling_matrix}
				\bsf{B}=\R{G}\left(\Z{G}\id{K}+\bsfZ{in}\right)^{-\mathrm{H}} \real{\bsfZ{in}} \left(\Z{G}\id{K}+\bsfZ{in}\right)^{-1}.
            \end{equation}

        \subsection{Input-Output Relation}
            The input-output relation is the expression that relates input voltages $\bsfvs{G}$ with output voltages $\bsfvs{L}$.
            Considering first a noiseless scenario (\ie $\bsfvs{N} = \bsfis{LNA} = \zeros{}$, $\bsfvs{L}=\bsfvs{R}$), it follows from \cref{eq:zin_zout} and Ohm's law that the physical channel matrix is given by
            \begin{equation}
                \label{eq:matrix_D}
                \begin{multlined}
                    \bsf{D}=\Z{L}\left(\Z{L}\id{N}+\bfZ{R}\right)^{-1}\bsfZ{RT}(\Z{G}\id{K}+\bsfZ{in})^{-1}.
                \end{multlined}
            \end{equation}
            Applying the Woodbury matrix identity~\cite{marti_asymptotic_2024},~\cite[Sec.~3.8]{meyer_matrix_2000} yields the following alternative expression for $\bsf{D}$:
            \begin{equation}
                \label{eq:matrix_D_2}
                \bsf{D} = \bsf{Q}\bsfZ{RT}(\Z{G}\id{K}+\bfZ{T})^{-1},
            \end{equation}
            where $\bsf{Q}$ is defined as
            \begin{equation}
                \label{eq:Q_matrix}
                \bsf{Q} = \Z{L}\left(\Z{L}\id{N}+\bsfZ{out}\right)^{-1}.
            \end{equation}

            The expression for the noise component follows from applying superposition and Thévenin's theorem at the load:
            \begin{equation}
                \label{eq:multiport_noise}
                \bsf{n} = \bsf{Q}\left(\bsfZ{out}\bsfis{LNA} - \bsfvs{N}\right).
            \end{equation}
            Its covariance matrix is given by
            \begin{equation}
                \label{eq:noise_cov}
                \cov{\bsf{n}} = \bsf{Q}\bsf{P}\bsf[H]{Q},
            \end{equation}
            with $\bsf{P}\! =\! \sigma_i^2(\R{N}^2\id{N}+\bsfZ{out}\bsfZ[H]{out}-2\R{N}\Re[\rho^*\bsfZ{out}]+\R{N}\Re[\bsfZ{out}])$.

            Letting $\bsf{y} = \bsfvs{L}$, $\bsf{z} = \bsf{n}$, and imposing $\power{T} = \expec{\bsf[H]{x}\bsf{x}}$ (up to a normalization constant) yields~\cite{ivrlac_toward_2010,damico_holographic_2024}:
            \begin{equation}
                \label{eq:it_model}
                \bsf{x} = \frac{\bsf[1/2]{B}\bsfvs{G}}{\sqrt{\R{G}}},\quad \bsf{H} = \sqrt{\R{G}}\bsf{D}\bsf[-1/2]{B},
            \end{equation}
            so that the information-theoretic channel coincides with the physical channel, up to a linear transformation.

        \subsection{Unilateral Approximation Discussion}
            \label{sec:ua_discussion}
            In the literature, the \acrlong{ua} corresponds to the assumption $\bsfZ{TR} \approx \zeros{K \times N}$, under which $\bsfZ{A}$ becomes the impedance matrix of a unilateral network~\cite[Sec.~11.2]{pozar_microwave_2012}, with $\bsfvs{T}\approx\bfZ{T}\bsfis{T}$ and $\bsfvs{R}\approx\bfZ{R}\bsfis{R}$, as follows from~\cref{eq:z_a}.
            Under this assumption, $\bsfZ{in} \approx \bfZ{T}$ and $\bsfZ{out} \approx \bfZ{R}$, so both the equivalent input and output impedances become deterministic.
            Consequently, it follows from \cref{eq:power_coupling_matrix,eq:matrix_D_2,eq:it_model} that, if $\bsfZ{RT}$ is a Gaussian random matrix, which is often justified in rich scattering, then $\bsf{H}$ corresponds to a correlated Rayleigh model.

            Of course, properly characterizing the unilateral approximation is fundamental in model-based receivers, such as those typically employed in high-frequency \acrshort{mimo}.
            Linear combiners, whether \acrfull{zf} or \acrfull{mmse}, require knowledge of the channel and the noise covariance.
            Thus, identifying the conditions under which fully coupled or simpler models can be considered is crucial.

            The unilateral model also offers computational advantages.
            Since $\bfZ{T}$ and $\bfZ{R}$ are static, they can be computed offline and stored, whereas in the fully coupled scenario they must be updated at every coherence interval.
            Moreover, because the equivalent input and output impedances are fixed, static matching networks can be designed and implemented, particularly in devices with a limited number of antennas.

            Finally, the unilateral approximation also allows users to be non-cooperative.
            In the general case, $\bsf{B}$ depends on $\bsfZ{RT}$ so each transmitting device would need to know the propagation channels of all other devices.
            Instead, under the unilateral approximation, only $\bfZ{T}$ is needed, which can be assumed diagonal for sufficiently separated users.
            Consequently, the power coupling matrix $\mathbf{B}$ also becomes diagonal, and the total power equation \cref{eq:power_T} decouples into $K$ independent per-user transmitted powers, allowing uncoordinated users to satisfy their own power budgets locally.

    \section{Problem Statement}
        \label{sec:problem_statement}

        The observations in the previous section highlight the importance of correctly identifying the regions where the unilateral approximation holds, since this determines when conventional channel models remain physically consistent and when their associated computational simplifications are valid.

        As previously stated, the input-side unilateral approximation assumes that $\bsfvs{T} \approx \bfZ{T}\bsfis{T}$.
        For this approximation to be valid, it must hold that $\norm{\bfDelta{in}\bfis{T}}{2} \ll \norm{\bfZ{T}\bfis{T}}{2}$ for every possible current, which motivates the following proposition.
        \begin{proposition}
            \label{prop:input-side}%
            Assuming that $\bfZ{T}$ is invertible, the condition
            \begin{equation}
                \label{eq:input-side}
                \sup_{\bfis{T} \neq \zeros{}}\frac{\norm{\bfDelta{in}\bfis{T}}{2}}{\norm{\bfZ{T}\bfis{T}}{2}} = \norm{\bfDelta{in}\bfZ[-1]{T}}{2} \ll 1,
            \end{equation}
            is necessary and sufficient for the input-side unilateral approximation to hold.
        \end{proposition}
        \begin{IEEEproof}
            Assume $\Vert\bfDelta{in}\bfZ[-1]{T}\Vert_2 \leq \varepsilon \ll 1$.
            Then,
            \begin{equation}
                \begin{aligned}
                    \norm{\bfDelta{in}\bfis{T}}{2} &= \norm{\bfDelta{in}\bfZ[-1]{T}\bfZ{T}\bfis{T}}{2} \! \leq \norm{\bfDelta{in}\bfZ[-1]{T}}{2} \norm{\bfZ{T}\bfis{T}}{2}\\
                    &\leq \varepsilon \norm{\bfZ{T}\bfis{T}}{2} \ll \norm{\bfZ{T}\bfis{T}}{2},
                \end{aligned}
            \end{equation}
            so~\cref{eq:input-side} is sufficient.
            Furthermore, if the condition $\norm{\bfDelta{in}\bfis{T}}{2} \ll \norm{\bfZ{T}\bsfis{T}}{2}$ must hold for every possible current $\bfis{T}$, it must hold for the worst-case current vector that maximizes the ratio in~\cref{eq:input-side}.
            Since this maximum ratio equals $\norm{\bfDelta{in}\bfZ[-1]{T}}{2}$, the spectral norm itself must be $\ll 1$.
        \end{IEEEproof}

        The condition in the preceding proposition is the one commonly used to characterize the unilateral approximation in the literature.
        Under this condition, the physical channel in \cref{eq:matrix_D} can be approximated by
        \begin{equation}
            \label{eq:d_unilateral}
            \bsf{D}_{\mathrm{UA}}= \underbrace{\Z{L}\left(\Z{L}\id{N}+\bfZ{R}\right)^{-1}}_{\mathbf{Q}_{\mathrm{UA}}}\bsfZ{RT}(\Z{G}\id{K}+\bfZ{T})^{-1}.
        \end{equation}
        However, \cref{prop:input-side} does not guarantee that $\bsfZ{out} \approx \bfZ{R}$.
        Consequently, it does not suffice to obtain the unilateral approximation of the noise term, which also requires the output-side condition stated in the following proposition.
        \begin{proposition}
            \label{prop:output-side}%
            Assuming that $\bfZ{R}$ is invertible, the condition
            \begin{equation}
            \label{eq:output-side}
                \sup_{\bfis{R} \neq \zeros{}}\frac{\norm{\bfDelta{out}\bfis{R}}{2}}{\norm{\bfZ{R}\bfis{R}}{2}} = \norm{\bfDelta{out}\bfZ[-1]{R}}{2} \ll 1,
            \end{equation}
            is necessary and sufficient for the output-side unilateral approximation to hold.
        \end{proposition}

        As with the input-side condition, \cref{prop:output-side} alone is insufficient to approximate all relevant quantities in the model, since the resulting power coupling matrix still depends on the full input impedance matrix.

        Based on the discussion above, we define the unilateral approximation as follows.
        \begin{definition}
            \label{def:unilateral_approximation}%
            In the context of a communication system, the unilateral approximation holds when \cref{prop:input-side,prop:output-side} are both satisfied.
        \end{definition}

        The main drawback of the conditions introduced above is their limited practical applicability.
        In particular, their calculation is not straightforward and is computationally complex, especially for large arrays, and they neither provide a quantitative measure of the approximation error nor specify how small the relevant norms must be for the approximation to hold.
        Thus, the main objective of this paper is to establish simpler conditions based on those presented herein which are suitable for large arrays or provide a bound on the maximum approximation error for each relevant variable.

        To do so, we consider a network comprising several single-antenna devices communicating with an \acrshort{ap} equipped with a large antenna array under \acrshort{los} propagation.
        Such a scenario is relevant to applications including short-range communications~\cite{bensky_short-range_2019} and wireless data centers~\cite{hamza_wireless_2016}, and also to high-frequency cellular communication systems~\cite{bacci_spherical_2023}, where the \acrshort{los} component is typically dominant.

        Beyond the circuit-theoretic analysis, we evaluate the impact of the unilateral approximation on \acrshort{se} by comparing the derived bounds with communication-theoretic performance, as equivalence at the circuit level does not directly translate to equivalent system-wide performance.

        \subsection{System Model}
            \label{sec:system_model}

            We consider the uplink of a multiuser \acrshort{mimo} system in which $K$ single-antenna devices each located at $(r_k, \theta_k, \varphi_k)$ communicate with an \acrshort{ap} equipped with an extremely large \acrfull{upa}.
            The array is located in the $yz$-plane and consists of $N$ antennas arranged in $2M+1$ rows, each containing $2M+1$ antennas, such that $N = (2M+1)\cdot(2M+1)$.
            Inter-element spacing is $d$, and the array spans a maximum dimension $D=2\sqrt{2} Md$, with corresponding Fresnel and Fraunhofer distances $\dist{\mathrm{r}}=0.62\sqrt{D^3/\lambda}$ and $\dist{\mathrm{F}}=2D^2/\lambda$.
            Each element in the array is assigned a pair of coordinates $(i_n,j_n)$ such that $i_n, j_n \in [-M, M]$.
            The index $1 \leq n \leq N$ uniquely determines an antenna, following a row-major ordering starting from $(i_1, j_1) \triangleq (-M, -M)$, which corresponds to the bottom left antenna.
            The distance between two dipoles is computed as
            \begin{equation}
                \label{eq:spacing_upa}
                d_{m,n}^{\mathrm{upa}} = d\sqrt{(i_m-i_n)^2+(j_m-j_n)^2},\ 1 \leq m,n \leq N.
            \end{equation}

            Each client operates under an individual transmit power constraint $P_{\mathrm{T},k} \le P_k$.
            For the purpose of this work, we assume uniform power allocation across users such that $P_{\mathrm{T},k} = 1/K$.
            The signal received at the \acrshort{ap} can be expressed in complex baseband notation as
            \begin{equation}
                \bsf{y} = \mathbf{H}\bsf{x} + \bsf{z},\quad \bsf{y}\in\cnumbers^N,
            \end{equation}
            where $\mathbf{H}\in\cnumbers^{N \times K}$ is the channel matrix; $\bsf{z}\in\cnumbers^N$ is additive noise with covariance $\cov{\bsf{z}} = \cov{\bsf{n}}$, up to a constant; and $\bsf{x}\in\mathcal{X}^{K}$ denotes the stacking of all the symbols transmitted by the clients.
            The corresponding expressions for these quantities are obtained from the multiport model through~\cref{eq:it_model}.

        \subsection{Receiver Structure}
            \label{sec:receiver_structure}

            As is common in high-frequency \acrshort{mimo} systems~\cite{marti_harnessing_2025,bacci_spherical_2023,marti_coherent_2025}, the receiver constructs the channel from a physical model using a set of previously estimated parameters.
            In the scenario considered in this work, the \acrshort{ap} requires knowledge of $\Z{G}$, $\Z{L}$, $\bfZ{R}$, $\bfZ{T}$, and $\bfZ{RT}$.
            Among these quantities, $\Z{G}$, $\Z{L}$, and $\bfZ{R}$ are static and can be calibrated offline, whereas $\bfZ{T}$ and $\bfZ{RT}$ depend on the locations of the client devices.
            Consequently, only the position of each device needs to be estimated.
            Since localization and parameter estimation are beyond the scope of this work, perfect knowledge of these quantities is assumed.

            To detect the information transmitted by the clients, the \acrshort{ap} employs a receive combiner vector $\mathbf{v}_k \in \cnumbers^{N}$ for the $k$-th client such that its \acrshort{se} is given by $\text{SE}_k = \log_2\left(1+\sinr{k}{}\right)$, with
            \begin{equation}
                \label{eq:effective_sinr}
                \sinr{k}{} = \frac{\left|\herm{\mathbf{v}_k}\mathbf{h}_k\right|^2}{\sum_{i=1}^{K}\left|\herm{\mathbf{v}_k}\mathbf{h}_i\right|^2 - \left|\herm{\mathbf{v}_k}\mathbf{h}_k\right|^2 + \herm{\mathbf{v}_k}\cov{\bsf{z}}\mathbf{v}_k}.
            \end{equation}
            The expression for $\mathbf{v}_k$ depends on whether the unilateral approximation is considered, and it is detailed in \cref{sec:spectral_efficiency}.

    \section{Mutual Coupling Impedance Matrices}
        \label{sec:mc_z_matrices}

        Mutual impedance between two elements is defined under open-circuit conditions as $Z_{21} = \frac{V_2}{I_1}|_{I_2=0}$.
        For thin antennas satisfying the \acrlong{cms} property~\cite{wasylkiwskyj_theory_1970}, open-circuited elements become invisible, justifying a pairwise computation of the impedance matrix entries.
        While intra-array coupling matrices $\bfZ{R}$ and $\bfZ{T}$ depend on the specific antenna geometry and are deferred to their respective sections, the inter-array coupling matrix $\bfZ{RT}$ can be treated as antenna-independent, up to a scaling factor.

        \subsection{\glsentryname{los} Propagation}
            \label{sec:los_prop}

            Assuming \acrshort{los} in the near field, the entries of the inter-array coupling matrix account for wavefront curvature via a spherical wavefront propagation model~\cite{lu_tutorial_2024}:
            \begin{equation}
                \label{eq:inter_array_hertzian}
                [\bfZ[]{RT}]_{n, k} = \Gamma_{k}\frac{\euler^{-\im\kappa r_{n,k}}}{\kappa r_{n,k}},\quad \begin{aligned}&1\leq n \leq N,\\ &1\leq k \leq K,\end{aligned}
            \end{equation}
            where $\Gamma_{k} \in \cnumbers$ denotes the aforementioned scaling factor in ohms, and $r_{n,k}$ is the distance between the $n$-th receive antenna and the $k$-th transmit antenna.
            Note that since all receiving antennas are identical and equally polarized, $\Gamma_k$ depends only on $k$ and not on $n$.

        \subsection{Electrical Length}
            \label{sec:electrical_length}

            Observe that the distances in \cref{eq:inter_array_hertzian} are normalized by the wavelength $\lambda$ through the wavenumber $\kappa$, yielding the electrical length~\cite[Ch.~11]{orfanidis_electromagnetic_2016}.
            Because all antenna dimensions (dipole length $l$, inter-element spacing $d$, and the array aperture $D$) are also defined proportionally to $\lambda$, and assuming frequency-invariant termination and noise parameters, the entire multiport model depends only on the electrical length $r/\lambda$.
            Therefore, when plotted against physical distance $r$ on a logarithmic scale, changing the carrier frequency simply shifts the curve horizontally by the corresponding wavelength ratio without altering its shape.

    \section{Error Bounds under Sufficient Conditions}
        \label{sec:error_bounds}

        We start our analysis by deriving sufficient conditions under which the unilateral approximation holds.
        Furthermore, we leverage them to characterize the approximation errors in $\bsf{D}$, $\bsf{B}$, and $\bsf{n}$ when the unilateral approximation is assumed.

        In particular, we show that these errors can be expressed in terms of the following sufficient conditions:
        \begin{align}
            \norm{\bfDelta{out}\bfZ[-1]{R}}{2} &\leq \frac{\norm{\bfDelta{out}}{2}}{\sigmamin(\bfZ{R})} = \varepsilon_\mathrm{out} \ll 1,\label{eq:suf_out}\\[1ex]
            \norm{\bfDelta{in}\bfZ[-1]{T}}{2} &\leq \frac{\norm{\bfDelta{in}}{2}}{\sigmamin(\bfZ{T})} = \varepsilon_\mathrm{in} \ll 1.\label{eq:suf_in}
        \end{align}

        These conditions provide closed-form error bounds and are independent of the antenna type under consideration.
        However, they become too conservative as $N \to \infty$ because the minimum singular value $\sigmamin(\bfZ{R})$ approaches zero in \acrshortpl{upa}.\footnote{For lossy antennas, where $\bfZ{R} \to \R{d}\id{N} + \bfZ{R}$, $\sigmamin(\bfZ{R})>0$ holds even as $N\to\infty$.}

        We next state the main results of this section as propositions.

        \begin{proposition}
            \label{prop:bound_D}%
            The error incurred by the unilateral approximation for the physical channel can be bounded either as
            \begin{equation}
                \frac{\norm{\mathbf{D}-\mathbf{D}_\mathrm{UA}}{2}}{\norm{\mathbf{D}_\mathrm{UA}}{2}} \leq \frac{\varepsilon_\mathrm{in}}{1-\varepsilon_\mathrm{in}},
            \end{equation}
            or as
            \begin{equation}
                \frac{\norm{\mathbf{D}-\mathbf{D}_\mathrm{UA}}{2}}{\norm{\mathbf{D}_\mathrm{UA}}{2}} \leq \frac{\varepsilon_\mathrm{out}}{1-\varepsilon_\mathrm{out}}.
            \end{equation}
        \end{proposition}
        \begin{IEEEproof}
            The proof leverages the Neumann series and standard bounds in perturbation theory.
            See~\cref{ap:proof_D}.
        \end{IEEEproof}

        Although $\mathbf{D}$ can be bounded by employing either the input or output conditions, the power coupling matrix can only be bounded through $\varepsilon_\mathrm{in}$.
        \begin{proposition}
            \label{prop:bound_B}%
            The error incurred by the unilateral approximation for the power coupling matrix can be bounded as
            \begin{equation}
                \frac{\norm{\mathbf{B}-\mathbf{B}_\mathrm{UA}}{2}}{\norm{\mathbf{B}_\mathrm{UA}}{2}} \leq \frac{(2+\mu_\mathrm{T}-\varepsilon_\mathrm{in})\varepsilon_\mathrm{in}}{(1-\varepsilon_\mathrm{in})^2},
            \end{equation}
            with $\mu_\mathrm{T} = \sigma_\mathrm{min}(\bfZ{T})/\lambda_\mathrm{min}(\Re[\bfZ{T}]) \geq 1$.
        \end{proposition}
        \begin{IEEEproof}
            See~\cref{ap:proof_B}.
        \end{IEEEproof}
        The parameter $\mu_\mathrm{T}$ characterizes the reactivity of the transmitting array, corresponding to $\mu_\mathrm{T}=1$ for matched transmitters.

        For noise, the mismatch is fully encapsulated in its covariance matrix, given in~\cref{eq:noise_cov}, and can be bounded only through $\varepsilon_\mathrm{out}$.
        \begin{proposition}
            \label{prop:bound_noise}%
            The error incurred by the unilateral approximation for the noise covariance matrix can be bounded as
            \begin{equation}
                \frac{\norm{\cov{\bsf{n}}-\cov{\bsf{n},\mathrm{UA}}}{2}}{\big\Vert\cov{\bsf{n},\mathrm{UA}}\big\Vert_2} \leq \frac{(2+\mu_\mathrm{N}-\varepsilon_\mathrm{out})\varepsilon_\mathrm{out}}{(1-\varepsilon_\mathrm{out})^2},
            \end{equation}
            with $\mu_\mathrm{N} = \frac{\sigmamin(\bfZ{R})(2\norm{\bfZ{R}}{2}+\R{N}(1+2|\rho|))}{\R{N}(\R{N}(1-|\rho|^2)+\lambda_\mathrm{min}(\Re[\bfZ{R}]))}$, up to $O(\varepsilon _{\text{out}}^{2})$ in the numerator.
        \end{proposition}
        \vspace{0.5ex}
        \begin{IEEEproof}
            See~\cref{ap:proof_noise}.
        \end{IEEEproof}
        Note that the parameter $\mu_\mathrm{N}$ is analogous to $\mu_\mathrm{T}$, accounting for the receiving array's mutual coupling, but also for the \acrshortpl{lna}' resistance and voltage-current coupling.

    \section{Necessary Conditions in Large Arrays}
        \label{sec:analysis_hertzian_dipoles}

        The analysis in the previous section provides bounds on the error incurred by the unilateral approximation, but it fails when the number of antennas becomes too large.
        Instead, we now propose a pair of necessary conditions for the unilateral approximation that can be computed in closed form as $N\to\infty$.
        In addition, this regime is of special interest since the large number of receiving antennas at the \acrshort{ap} is more likely to induce non-negligible currents at the users.

        To derive these necessary conditions, we lower-bound \cref{eq:input-side,eq:output-side} employing the Frobenius norm.
        In particular,
        \begin{equation}
            \norm{\bfDelta{in}}{F} = \norm{\bfDelta{in}\bfZ[-1]{T}\bfZ{T}}{F} \leq \norm{\bfDelta{in}\bfZ[-1]{T}}{2} \norm{\bfZ{T}}{F},
        \end{equation}
        so, dividing by $\norm{\bfZ{T}}{F}$, the input-side condition in \cref{eq:input-side} can be lower bounded as
        \begin{equation}
            \label{eq:condition_input}
            \frac{\norm{\bfDelta{in}}{F}}{\norm{\bfZ{T}}{F}} \leq \norm{\bfDelta{in}\bfZ[-1]{T}}{2} \ll 1.
        \end{equation}
        The same rationale applies to the output-side condition:
        \begin{equation}
            \label{eq:condition_output}
            \frac{\norm{\bfDelta{out}}{F}}{\norm{\bfZ{R}}{F}} \leq \norm{\bfDelta{out}\bfZ[-1]{R}}{2} \ll 1.
        \end{equation}

        These two conditions correspond to those studied in~\cite{marti_asymptotic_2024} for an \acrshort{ap} equipped with a \acrshort{ula} communicating with a single user.
        In this section, we extend that analysis to \acrshortpl{upa} in a multiuser scenario.

        \subsection{Mutual Impedance Matrix of Hertzian Dipoles}
            \label{sec:impedance_hertzian_dipoles}

            Hertzian dipoles are the simplest antennas that satisfy Maxwell's equations.
            Although they are not physically realizable, they admit analytical expressions for the mutual impedance, providing a convenient parameter-independent framework for communication-theoretic analysis.
            In addition, they are utilized as building blocks of more complex geometries~\cite[Sec.~4.2]{balanis_antenna_2016}.

            As mentioned above, mutual impedance depends on the relative positions of antenna elements.
            In the case of a \acrshort{upa} of Hertzian dipoles, three configurations arise: side-by-side, collinear, and parallel-in-echelon dipoles~\cite[Sec.~8.6]{balanis_antenna_2016}.
            The intra-array coupling impedance matrix $\bfZ{R}$ is then constructed by applying the mutual impedance expression derived in \cref{ap:mutual_z_hertzian} to each antenna pair.
            Assuming that loading coils are inserted between each antenna and its feedline to compensate for the capacitive reactance of the dipoles~\cite[Ch.~2]{pozar_microwave_2012}, the \acrshort{ap} intra-array coupling impedance matrix is given~by
            \begin{equation}
            	\label{eq:coupling_hertzian_dipoles_bs}
                [\bfZ{R}]_{m,n} =
                \begin{cases}
            		\R{r}, &m=n,\\
            		\frac{3}{2}\R{r}\,\im\,\euler^{-\im\kappa \dist[upa]{m,n}}\psi(\dist[upa]{m,n}, \theta_{m,n}), &m\neq n,
            	\end{cases}
            \end{equation}
            where $\theta_{m,n} = \arccos{\frac{d\cdot|j_m-j_n|}{\dist[upa]{m,n}}}$ is the angle between the $m$-th and $n$-th antenna elements, and $\psi(r,\theta)$ is first defined in~\cref{eq:psi_hertzian} in \cref{ap:mutual_z_hertzian}.
            In~\cref{eq:coupling_hertzian_dipoles_bs}, vertical dipoles aligned along the $z$-axis are assumed.

            Because the client devices locations are arbitrary, the impedance matrix $\bfZ{T}$ is computed numerically from \cref{eq:z_hertzian_general} when needed.

        \subsection{Output-Side Asymptotic Analysis}
            \label{sec:los_multiuser}
            We begin the analysis by studying the output-side condition given in \cref{eq:condition_output}.

            Taking into account that the Frobenius norm is submultiplicative~\cite[Sec.~5.2]{meyer_matrix_2000}, we can upper bound the \acrfull{lhs} in \cref{eq:condition_output} as
            \begin{equation}
                \label{eq:bound_lhs_ul}
                \begin{aligned}
                    &\norm{\bfZ{RT}(\Z{G}\id{K} + \bfZ{T})^{-1}\bfZ{TR}}{F}\\
                    &\qquad\leq\norm{\bfZ{RT}}{F}^2\cdot\norm{(\Z{G}\id{K} + \bfZ{T})^{-1}}{F} \ll \norm{\bfZ{R}}{F},
                \end{aligned}
            \end{equation}
            which implies \cref{eq:condition_output}.
            The main advantage of \cref{eq:bound_lhs_ul} is that the users impedance matrix does not depend on $N$.
            Therefore, it suffices to show that $\norm{\bfZ{R}}{F}$ grows faster than $\norm{\bfZ{RT}}{F}^2$ asymptotically.

            Since the array is assumed to be composed of Hertzian dipoles, we can apply~\cref{eq:bound_lhs_ul} using $\bfZ{R}$ from~\cref{eq:coupling_hertzian_dipoles_bs} and $\bfZ{RT}$ from~\cref{eq:inter_array_hertzian}.

            We begin by computing the norm of the \acrshort{ap} coupling matrix:
            \begin{equation}
                \label{eq:frob_bs}
                \begin{aligned}
                    \hspace{-0.5ex}\norm{\bfZ{R}}{F} &= \sqrt{\sum_{m,n=1}^{N}|[\bfZ{R}]_{m,n}|^2} = \sqrt{\sum_{m=1}^{N}\norm{[\bfZ{R}]_{m, *}}{2}^2}\\
                    &= \sqrt{N}\cdot\norm{[\bfZ{R}]_{\bolddot, *}}{2}.
                \end{aligned}
            \end{equation}
            The last step follows from the assumption of an infinite array.
            Indeed, since the $n$-th row of $\bfZ{R}$ represents the mutual coupling between the $n$-th antenna and all the others, each antenna in an infinite array has the same surrounding environment.
            Consequently, all rows have identical Euclidean norms.

            In order to compute the Euclidean norm of $[\bfZ{R}]_{\bolddot, *}$, it is more convenient to split it into four parts: the self-impedance, the parallel-in-echelon terms, the $M$ terms for the side-by-side dipoles and the $M$ terms for collinear dipoles.
            Using the symmetry of the square grid, the norm can be expressed as
            \begin{equation}
                \label{eq:upa_coupling_norm}
                \begin{aligned}
                    &\norm{[\bfZ{R}]_{\bolddot, *}}{2}^2 = \R{r}^2 + 9\R{r}^2\smashoperator{\sum_{i,j=1}^{M}}\ \left|\psi\left(d\sqrt{i^2+j^2},\vartheta_{i,j}\right)\right|^2\\ & + \frac{9}{2}\R{r}^2\smashoperator{\sum_{i=1}^{M}}\ \left|\psi\left(d\cdot i, \frac{\pi}{2}\right)\right|^2
                    + \frac{9}{2}\R{r}^2\smashoperator{\sum_{j=1}^{M}}\,\left|\psi\left(d\cdot j, 0\right)\right|^2,
                \end{aligned}
            \end{equation}
            with $\vartheta_{i,j}=\arcsin{\frac{i}{\sqrt{i^2+j^2}}}$.
            The coupling function $\psi$ is defined in~\cref{eq:psi_hertzian}, with squared absolute value:
            {
            \small
            \begin{equation}
                \label{eq:psi_squared}
                |\psi(r, \theta)|^2 = \frac{\sin^4\!\theta}{\kappa^2 r^2} - \frac{1{-}2\cos^2\!\theta{-}3\cos^4\!\theta}{\kappa^4 r^4} + \frac{(1{-}3\cos^2\!\theta)^2}{\kappa^6 r^6}.
            \end{equation}
            }

            The last two sums in~\cref{eq:upa_coupling_norm} are equivalent to those obtained for \acrshortpl{ula} of side-by-side and collinear dipoles.
            Since the former is known to converge~\cite{marti_asymptotic_2024}, and the latter is upper bounded by it because collinear dipoles exhibit weaker mutual coupling (see~\cref{ap:mutual_z_hertzian}), these sums are convergent, and
            it remains only to analyze the divergence of the parallel-in-echelon dipole contribution.
            Putting the explicit value of $|\psi|$ in this sum gives
            \begin{equation}
                \frac{1}{\kappa^2 d^2} \sum_{i,j=1}^{M} \frac{i^4}{(i^2+j^2)^3} + \text{higher order terms}.
            \end{equation}
            These higher order terms lead to convergent series, as follows from \cref{lem:sum_no_const}.
            In contrast, \cref{lem:double_series_2} shows that the first term diverges as
            \begin{equation}
                \frac{1}{\kappa^2 d^2} \frac{3\pi}{16} \ln M
            \end{equation}
            when $M \to \infty$.
            Consequently,
            \begin{equation}
                \label{eq:ap_norm_order}
                \norm{[\bfZ{R}]_{\bolddot, *}}{2}^2 \sim \frac{27 \pi}{32} \frac{\R{r}^2}{\kappa^2 d^2} \ln N
            \end{equation}
            as $N \to \infty$.
            Finally, substituting the square root of \cref{eq:ap_norm_order} into~\cref{eq:frob_bs} yields
            \begin{equation}
                \norm{\bfZ{R}}{F} \sim \frac{3\sqrt{6\pi}}{8} \frac{\R{r}}{\,\kappa d\,} \sqrt{N \ln N }.
            \end{equation}

            To complete the output-side analysis, we compute the norm of the reflected coupling matrix.
            To this end, we particularize the model in \cref{eq:inter_array_hertzian} to our setup, yielding $\Gamma_{k} = \num{1.5}\im\R{r}$ for parallel Hertzian dipoles~\cite{marti_asymptotic_2024}.
            Furthermore, the distance from the $k$-th device to the $n$-th array element is given by
            \begin{equation}
                \label{eq:distance_los}
                r_{n,k}^{2}=x_k^2+d^2\left(\left(i_n-\frac{y_k}{d}\right)^2+\left(j_n-\frac{z_k}{d}\right)^2\right),
            \end{equation}
            where $(x_k,y_k,z_k)$ is the device location in Cartesian coordinates.
            When $N \to \infty$, the finite offset $(y_k/d, z_k/d)$ does not affect the asymptotic behavior of $\norm{\bfZ{RT}}{F}$, so \cref{eq:distance_los} can be replaced by
            \begin{equation}
                r_{n,k}^2 = x_k^2+d^2(i_n^2+j_n^2).
            \end{equation}

            Letting $\mathcal{L} = [-M, M] \times [-M, M] \subset \znumbers^2$, the Frobenius norm of the inter-array coupling matrix can be expressed as
            \begin{equation}
                \label{eq:norm_zrt_los}
                \norm{\bfZ{RT}}{F}^2 = \frac{9}{4}\frac{\R{r}^2}{\kappa^2}\sum_{k=1}^{K}\sum_{(i,j)\in\mathcal{L}} \frac{1}{x_k^2 + d^2(i^2+j^2)}.
            \end{equation}
            Since $M \to \infty$, it follows from \cref{lem:double_series_1} that the series in~\cref{eq:norm_zrt_los} diverges as
            \begin{equation}
                \norm{\bfZ{RT}}{F}^2 \sim \frac{9\pi\R{r}^2}{4 d^2\kappa^2} K \ln N.
            \end{equation}

            Putting everything together, since the \acrshort{lhs} of \cref{eq:bound_lhs_ul}, which upper-bounds the reflected coupling, diverges as $\ln N$, whereas the intra-array coupling diverges as $\sqrt{N\ln N}$, we conclude that, under the stated assumptions, the necessary output-side condition for the unilateral approximation is not violated asymptotically from a circuit-theoretic perspective.

        \subsection{Input-Side Asymptotic Analysis}
            \label{sec:los_multiuser_input}

            The analysis of the input-side condition in \cref{eq:condition_input} is not so straightforward, since the \acrshort{lhs} directly depends on the inverse of the \acrshort{ap} impedance matrix.
            Nevertheless, this same matrix is block-Toeplitz with Toeplitz blocks.
            In particular, $\bfZ{R} \in \mathcal{BTTB}_{\sqrt{N},\sqrt{N}}$, so it diagonalizes asymptotically with the $\sqrt{N}\times\sqrt{N}$ 2D Fourier transform~\cite{gray_toeplitz_2005}, \cite[Sec.~5.8]{davis_circulant_1986}:
            \begin{equation}
                \label{eq:zap_fourier}
                \bfZ{R} = \herm{\mathbf{U}} \mathbf{\Lambda} \mathbf{U},\quad \mathbf{U} = \mathbf{F}_{\sqrt{N}} \otimes \mathbf{F}_{\sqrt{N}},
            \end{equation}
            where $\mathbf{\Lambda}$ is a diagonal matrix containing the eigenvalues $\{\lambda_n\}_{1\leq n\leq N}$  of $\bfZ{R}$ as $N\to\infty$.
            For simplicity of presentation, we shall order the eigenvalues in a 2D grid such that $\lambda_n = \lambda_{p,q}$ with $n = (p + M)(2M+1) + (q + M + 1)$, for $p,q \in \{-M, \dots, M\}$.

            Substituting \cref{eq:zap_fourier} into \cref{eq:condition_input} yields the following expression for the $k,l$-th element in the matrix:
            \begin{equation}
                \label{eq:akl_1}
                A_{k,l} = [\bfDelta{in}]_{k,l} = \trans{(\mathbf{U}^* \mathbf{z}_{\text{RT},k})}(\Z{L}\id{N}+\mathbf{\Lambda})^{-1}(\mathbf{U} \mathbf{z}_{\text{RT},l}),
            \end{equation}
            with $\mathbf{z}_{\text{RT},k}$ the $k$-th column of $\bfZ{RT}$.
            Leveraging the spatial frequency representation from \cref{ap:spectral_representation},
            \begin{equation}
                \label{eq:akl_approx}
                A_{k,l} \approx \frac{1}{N d^4}\sum_{p,q = -M}^{M} \frac{\tilde{z}_{\text{RT},k}(-k_{y}, -k_{z})\tilde{z}_{\text{RT},l}(k_{y}, k_{z})}{\Z{L}+\lambda(k_y, k_z)},
            \end{equation}
            where $k_y = p\cdot\frac{2\pi}{d\sqrt{N}}$ and $k_z = q\cdot\frac{2\pi}{d\sqrt{N}}$ are the wavenumber components along the $y$- and $z$-axes, with $\sqrt{N}=2M+1$.

            As $N \to \infty$, the sum in~\cref{eq:akl_approx} converges to the integral
            \begin{equation}
                \label{eq:I_kl}
                I_{k,l} = \frac{-\Gamma_k\Gamma_l}{d^2\kappa^2}\iint_{\mathcal{BZ}} \frac{\dd k_y \dd k_z \euler^{-\im(k_x\nabla x + k_y\Delta y + k_z\Delta z)}}{k_x^2 \cdot (\Z{L}+\lambda(k_y,k_z))},
            \end{equation}
            where $\mathcal{BZ} = \left[-\frac{\pi}{d}, \frac{\pi}{d}\right] \times \left[-\frac{\pi}{d}, \frac{\pi}{d}\right]$ denotes the first Brillouin Zone, $\nabla x=x_l+x_k$, $\Delta y=y_l-y_k$, and $\Delta z =z_l-z_k$.
            While the $1/k_x$ divergence of $\lambda(k_y, k_z)$ compensates for the $1/k_x^2$ singularity in~\cref{eq:I_kl} at generic boundary points (see $\hat{\lambda}(k_y, k_z)$ in \cref{ap:spectral_representation}), this compensation fails at the end-fire poles $(0, \pm \kappa)$ where $\kappa^2 - k_z^2 = 0$.
            However, transforming to coordinates $(u, k_y)$ with $u = \kappa - \sqrt{k_y^2 + k_z^2}$, we have $k_x \approx \sqrt{2\kappa u}$ and $\kappa^2 - k_z^2 \approx 2\kappa u + k_y^2$ as $u \to 0$.
            Thus, the denominator from~\cref{eq:I_kl} scales as $2\kappa u\Z{L} + \frac{3\pi\R{r}}{\kappa^3}\sqrt{2\kappa u}(2\kappa u+k_y^2)$, and integrating over $k_y$ yields a $u^{-3/4}$ singularity, which is integrable over $u$.

            Given that the Frobenius norm requires the absolute value of each matrix term,
            \begin{equation}
                |A_{k,l}| \approx \frac{|\Gamma_k \Gamma_l|}{4\pi^2}\left(\frac{\lambda}{d}\right)^2 |C_{k,l}|,
            \end{equation}
            with $C_{k,l}$ the double integral in~\cref{eq:I_kl}.

            In conclusion, $\bfZ[T]{RT}(\Z{L}\id{N}+\bfZ{R})^{-1}\bfZ{RT}$ is a $K \times K$ matrix whose elements are all bounded.
            Thus, unlike the output-side condition, the input-side condition is independent of $N$.

    \section{Numerical Analysis}
        \label{sec:numerical_analysis}

        The preceding results show that \textit{(i)} the errors introduced by applying the unilateral approximation to the matrices of the multiport model remain bounded by the sufficient conditions in~\cref{eq:suf_out,eq:suf_in}, and \textit{(ii)} scaling the antenna count and physical aperture does not inherently violate the unilateral approximation.
        Below, we evaluate the non-asymptotic behavior of these conditions and assess the accuracy of these bounds.

        To this end, we consider Hertzian dipoles of length $l=\lambda/20$ operating at $f=\qty{30}{\GHz}$, yielding a radiation resistance $\R{r} \approx \qty{2}{\ohm}$.
        To prevent generator and load impedances from provoking a regularization that could hide the effects of mutual coupling, we set them to a value of the same order of magnitude as $\R{r}$.
        In particular, $\Z{L}=\Z{G} = \qty{5}{\ohm}$.
        The remaining parameters are $B_{\mathrm{W}} = \qty{20}{\MHz}$, $T_{\mathrm{A}}= \qty{290}{\K}$, $\R{N} = \qty{5}{\ohm}$, $d=\lambda/2$, and $\rho=0.1$.
        Inter-array coupling is computed using the exact expression in~\cref{eq:z_hertzian_general}, rather than the point-source approximation adopted in the asymptotic analysis, in order to provide meaningful results in the reactive near field.\footnote{Employing \cref{eq:z_hertzian_general} for $\bfZ{RT}$ requires the dipoles at the \acrshort{ap} and the client devices to be parallel.}

\begin{figure}[t]
            \centering
            \includegraphics[width=\columnwidth]{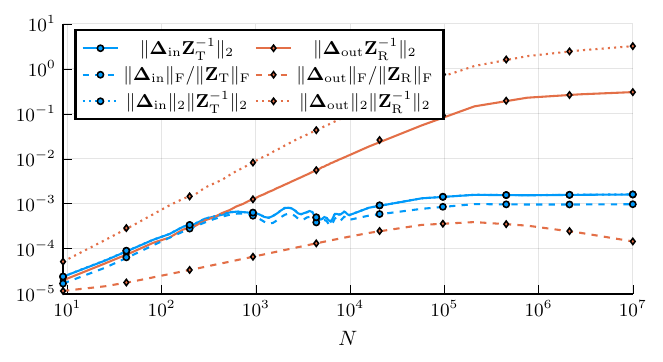}
            \vspace{-4.5ex}
            \caption{Conditions for the unilateral approximation as $N$ increases, with $K=5$ devices deployed \qty{1}{\m} from the \acrshort{ap}, and $d=\lambda/2$ at $f=\qty{30}{\GHz}$.}
            \label{fig:los_mimo_coupling_hertzian}
            \vspace{-2ex}
        \end{figure}

        In \cref{fig:los_mimo_coupling_hertzian}, we depict the defining conditions for the unilateral approximation (solid) given in \cref{eq:input-side,eq:output-side}, together with the sufficient (dotted) and  necessary (dashed) conditions from \cref{eq:suf_out,eq:suf_in}, and \cref{eq:condition_input,eq:condition_output}.
        A total of $K=5$ ground users deployed uniformly across a \qty{120}{\degree} azimuthal arc \qty{1}{\m} from the \acrshort{ap} are considered.
        We observe that the input-side conditions are very tight in all three cases.
        Indeed, the sufficient and defining conditions overlap completely.
        The reason is that at a \qty{1}{\m} separation, the inter-user transmit matrix $\bfZ{T}$ is nearly a scaled identity matrix, so multiplicativity yields an equality.
        The gap with the necessary condition is noticeable and would increase with $K$.
        Nevertheless, all three cases remain bounded around $10^{-3}$ in the asymptotic regime, suggesting that the unilateral approximation holds.
        Regarding the output side, the gap among all three conditions is considerable.
        Although the defining and sufficient conditions follow the same behavior, the necessary condition severely underestimates the requirements for the approximation to hold.
        Moreover, for a very large number of antennas, the defining ratio approaches one, indicating that the unilateral approximation does not hold.
        Yet, this occurs beyond any practical or even reasonable number of antennas.

\begin{figure}[t]
            \centering
            \includegraphics[width=\columnwidth]{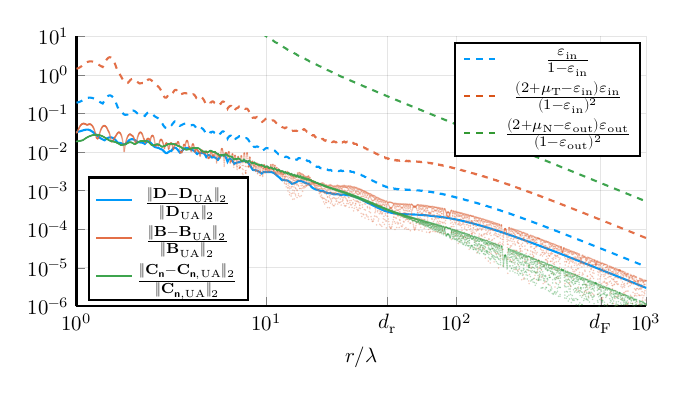}
            \vspace{-5.5ex}
            \caption{Relative errors and theoretical bounds under the unilateral approximation in $\mathbf{D}$, $\mathbf{B}$, and $\cov{\bsf{n}}$ versus electrical distance between the \acrshort{ap} and $K=5$ users, assuming a $25 \times 25$ \acrshort{upa} with $d=\lambda/2$ at $f=\qty{30}{\GHz}$. The Fresnel and Fraunhofer distances are denoted by $\dist{\mathrm{r}}$ and $\dist{\mathrm{F}}$.}
            \label{fig:los_mimo_bounds_hertzian}
            \vspace{-1.5ex}
        \end{figure}

        On the other hand, \cref{fig:los_mimo_bounds_hertzian} illustrates the approximation errors in $\mathbf{D}$, $\mathbf{B}$, and $\cov{\bsf{n}}$ versus distance for $K=5$ users communicating with a $25 \times 25$ \acrshort{upa}, alongside the analytical bounds (dashed lines) derived in \cref{sec:error_bounds}.
        As expected, the approximation error decreases with increasing distance.
        Notably, all three matrices exhibit almost identical error values, which are closely tracked by the input-side bound in the case of $\mathbf{D}$ and $\mathbf{B}$.
        In contrast, the output-side bound shows a noticeable discrepancy relative to the exact value and diverges for distances below $10\lambda$.
        This reflects the fact that due to a small $\sigmamin(\bfZ{R})$, the growth of $\norm{\bfDelta{out}}{2}$ at short distances drives $\varepsilon_{\text{out}} \to 1$.
        Consequently, the denominator $1 - \varepsilon_{\text{out}}$ vanishes, causing the analytical bound to diverge as the Neumann series convergence threshold ($\varepsilon_{\text{out}} < 1$) is approached.

        Overall, numerical results validate our theoretical analysis, showing that the unilateral approximation requires caution at short ranges or for extremely large arrays.

    \section{Spectral Efficiency}
        \label{sec:spectral_efficiency}

        Up to this point, we have exclusively focused on a circuit-theoretic validation of the unilateral approximation.
        However, such an analysis does not directly quantify its impact on communication performance.
        To this end, we consider a \acrshort{zf} combiner~\cite[Sec.~4.1]{bjornson_massive_2017},
        \begin{equation}
            \mathbf{V}_{\mathrm{ZF}} = \mathbf{H}\left(\herm{\mathbf{H}}\mathbf{H}\right)^{-1} \in \cnumbers^{N \times K},
        \end{equation}
        so $\mathbf{v}_k$ in~\cref{eq:effective_sinr} corresponds to the $k$-th column of $\mathbf{V}_{\mathrm{ZF}}$.
        Since the \acrshort{zf} combining matrix depends only on the channel matrix, it isolates the impact of channel mismatch introduced by the unilateral approximation from other receiver design choices.

        Specifically, we compare two \acrshort{zf} combiners.
        The first is the matched combiner, constructed from the exact channel model derived in~\cref{eq:it_model}, and denoted by $\mathbf{V}_{\mathrm{ZF}}$.
        The second is a mismatched combiner obtained from the same model under the unilateral approximation ($\bsfZ{in}=\bfZ{T}$ and $\bsfZ{out}=\bfZ{R}$), which we denote by $\mathbf{V}_{\mathrm{ZF}}^{\mathrm{UA}}$.
        This comparison allows us to quantify the impact of the unilateral approximation on the \acrshort{se}.

        The considered setup is the same as that in \cref{sec:numerical_analysis}.
        That is, $K=5$ users communicate with a $25\times25$ \acrshort{upa}.
        In \cref{sec:se_hertzian}, the array elements are assumed to be Hertzian dipoles, whereas in \cref{sec:hw_dipoles}, we consider half-wavelength dipoles in order to verify our results with realistic antennas.

        \subsection{Hertzian Dipoles}
            \label{sec:se_hertzian}

\begin{figure}
                \centering
                \includegraphics[width=\columnwidth]{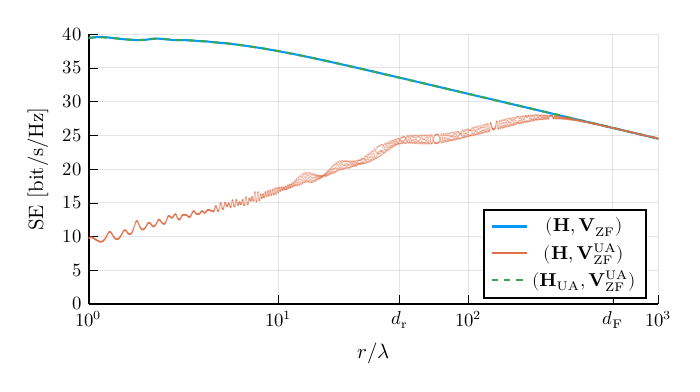}
                \vspace{-5.5ex}
                \caption{Spectral efficiency with respect to the electrical distance for $K=5$ and a $25 \times 25$ \acrshort{upa} with $d=\lambda/2$ at $f=\qty{30}{\GHz}$. The Fresnel and Fraunhofer distances are denoted by $\dist{\mathrm{r}}$ and $\dist{\mathrm{F}}$.}
                \label{fig:se_los_r}
            \end{figure}

            In \cref{fig:se_los_r}, we depict the average \acrshort{se} using either the matched \acrshort{zf} combiner, $\mathbf{V}_{\mathrm{ZF}}$, or the mismatched \acrshort{zf} combiner constructed under the unilateral approximation, $\mathbf{V}_{\mathrm{ZF}}^{\mathrm{UA}}$.
            For reference, an artificial unilateral system is also included.
            It can be observed that the coupling-aware combiner achieves essentially the same \acrshort{se} as the unilateral scenario.

            In contrast, the mismatched \acrshort{zf} combiner yields a substantially lower \acrshort{se} and exhibits oscillatory behavior in the near field.
            Contrary to what one might expect, these oscillations are not caused by wavefront curvature, but by the channel mismatch itself, since the rapidly varying receiver-to-transmitter coupling contributions to the input and output impedances are neglected in the mismatched \acrshort{zf} design.
            As the distance increases, however, the reflected coupling decreases and eventually becomes negligible.
            Indeed, this occurs when the approximation errors from \cref{fig:los_mimo_bounds_hertzian} are between $10^{-4}$ and $10^{-5}$.

            The observed \acrshort{se} degradation, despite an almost negligible circuit-theoretic mismatch, stems from the inherent structure of the \acrshort{zf} combiner: a small deviation in the estimated interference subspace results in an amplified residual interference term.
            A \acrshort{mmse} estimator would likely suffer from a similar degradation, since the channel and noise covariance matrices are also affected by the model mismatch~\cite{bacci_spherical_2023}.

        \subsection{Half-Wavelength Dipoles}
            \label{sec:hw_dipoles}

            Hertzian dipoles have allowed us to develop an analytical framework to assess the unilateral approximation when employing large arrays.
            However, Hertzian dipoles are not physically realizable.
            For this reason, in the sequel we consider arrays equipped with half-wavelength dipoles, a practical antenna implementation.

            The first step in analyzing the performance of half-wavelength dipole arrays is to determine their impedance matrices.
            An accurate approximation for center-fed half-wavelength dipoles is the sinusoidal current model \cite{damico_holographic_2024}, \cite[Sec.~24.4]{orfanidis_electromagnetic_2016}, which yields closed-form expressions for the mutual impedance in the parallel-in-echelon configuration.
            Following the notation in \cref{sec:impedance_hertzian_dipoles},
            \begin{equation}
                \label{eq:hw_impedance_matrix}
                [\mathbf{Z}_{\boldsymbol{\cdot}}]_{m,n} = \begin{cases}
                    \R{r}+\im\X{A}, &m=n,\\
                    \zeta(d_{m,n}, \theta_{m,n}), &m \neq n,
                \end{cases}
            \end{equation}
            where $\boldsymbol{\cdot}$ denotes either $\mathrm{R}$ or $\mathrm{T}$.
            The inter-array impedance matrix $\bfZ{RT}$ is obtained by evaluating the same mutual impedance function $\zeta(d,\theta)$ for every \acrshort{ap}--device antenna pair.

            In~\cref{eq:hw_impedance_matrix}, the diagonal terms represent the self-impedance of a half-wavelength dipole, which is known to be $73+\im42.5\,\Omega$.
            The mutual impedance between two parallel-in-echelon dipoles, denoted by $\zeta(d,\theta)$, depends on the relative distance $d_{m,n}$ and orientation $\theta_{m,n}$ between each pair of antennas.
            Since the corresponding expression is rather cumbersome, we refer the reader to~\cite[Eq.~(8-70)]{balanis_antenna_2016}.

\begin{figure}
                \centering
                \includegraphics[width=\columnwidth]{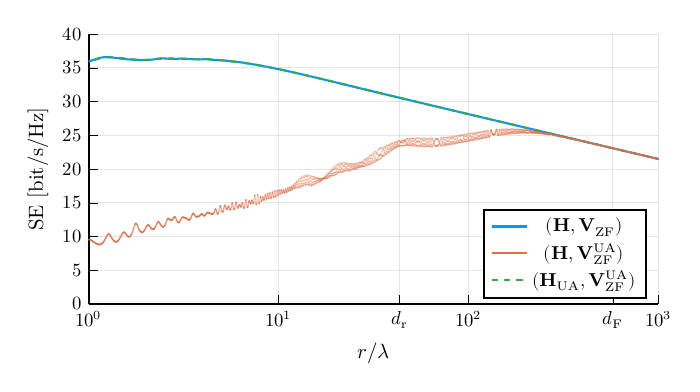}
                \vspace{-4.5ex}
                \caption{Spectral efficiency with respect to the electrical distance for $K=5$ and a $25 \times 25$ \acrshort{upa} equipped with half-wavelength dipoles separated $d=\lambda/2$. The Fresnel and Fraunhofer distances are denoted by $\dist{\mathrm{r}}$ and $\dist{\mathrm{F}}$.}
                \label{fig:se_los_r_hw}
            \end{figure}

            In \cref{fig:se_los_r_hw}, the \acrshort{se} of the same system as before is plotted, but with $\Z{L} = \Z{G} = 186-\im31.6\,\Omega$ as in~\cite{lehmeyer_lna_2015}, since the self-impedance is now $73+\im42.5\,\Omega$.
            As in the Hertzian dipole case, the receiver designed under the unilateral approximation suffers significant performance degradation at short distances but converges to the matched receiver inside the radiative near field.
            Indeed, both plots are almost identical, reinforcing the idea that Hertzian dipoles can be employed for system analysis.

    \section{Conclusions}
        In this paper, we investigated the validity of the unilateral approximation in communication systems from both circuit- and communication-theoretic perspectives.
        First, we discussed the state-of-the-art definition of the approximation and argued that not only the input side but also the output side must be taken into account.
        From these representations, we established sufficient conditions that yield analytical error bounds on the core matrices characterizing the system; specifically, the channel, noise covariance, and power coupling matrices.
        Furthermore, we derived necessary conditions and evaluated them analytically for asymptotically large arrays, proving that scaling the antenna count and physical aperture does not inherently violate the unilateral assumption.

        We also investigated the impact of the unilateral approximation on communication performance.
        Specifically, we showed that in multiuser scenarios, practical linear receivers (such as~\acrshort{zf}) designed under the unilateral assumption suffer substantial \acrshort{se} degradation at short distances due to model mismatch.
        In particular, we demonstrated that systems employing either Hertzian or half-wavelength dipoles suffer from the same penalty and can achieve similar spectral efficiencies, provided that the generator and load impedances are of the same order of magnitude as the antenna self-impedance.
        These results indicate that, although the unilateral approximation may be valid from a circuit-theoretic perspective, receiver algorithms must still account for receiver-to-transmitter mutual coupling to avoid significant performance degradation at short distances.

        Some extensions of this work remain open.
        A limitation of this study is the assumption that all antennas are aligned along the $z$-axis, which enables the use of closed-form expressions for half-wavelength dipoles.
        Although we do not expect different polarization configurations to alter the conclusions, this should be formally investigated.
        Furthermore, other antenna geometries, such as patch antennas, could also be considered.
        Finally, the design of matching networks specifically aimed at compensating for receiver-to-transmitter coupling and improving the performance of mismatched receivers represents an interesting direction for future research.

    \appendix
        \subsection{Divergence of Sums over Two-Dimensional Lattices}
            \label[appendix]{ap:lattice_sum}

            Throughout the paper we need to study the divergence of a number of double series.
            For positive and decreasing ordinary series this can be performed via the integral test, which compares the series with the integral of a continuous decreasing function (see for instance \cite[Sec.~9.2.6]{bartle_introduction_2011});
            it concludes that the series is convergent or divergent if and only if the corresponding integral also is.
            In the convergent case it also gives an estimation of the series in terms of the integral.
            However, we are more interested in the order of divergence.
            \begin{lemma}
                \label{lem:series}%
                Let $L < M$ be integers, $f \colon [L,M] \to \rnumbers$ a continuous and positive function, and $a_n=f(n)$.
                Consider the series and the integral
                \begin{equation}
                    S = \sum_{n=L}^{M} a_n,\quad I = \int_L^{M} \!\! f(x) \,\dd x.
                \end{equation}
                Let $K = \max_{x \in [L,M]} f(x)$.
                Then:
                \begin{enumerate}
                    \item If $f$ is monotonic, then $K = \max(a_L,a_M)$, and $|I-S| \leq K$.
                    \item If $f$ is increasing on $[L,N]$ and decreasing on $[N,M]$, then $K = f(N)$, and $|I-S| \leq 2K$.
                \end{enumerate}
            \end{lemma}
            \begin{corollary}
                With the preceding notations, we suppose that $f$ is defined on the interval $[L,+\infty)$ and is decreasing beyond a certain point.
                Then the series $S_M = \sum_{n=L}^{M} a_n$ and the integral $I_M = \int_L^{M} \!\! f(x) \,\dd x$ are either both convergent or both divergent as $M \to \infty$.
                In the latter case, they are asymptotically equivalent, that is, $S_M \sim I_M$.
            \end{corollary}

            As remarkable examples, it is well known that $\sum_{n=1}^\infty \frac 1{n^2} = \pi^2/6$, whereas $\sum_{n=1}^M \frac 1n \sim \ln M$.
            In what follows we will make use of the lemma, as well as the idea behind the corollary.

            \medskip
            \noindent\textit{\ref{ap:lattice_sum}.1}
            The first sum we consider is the double series
            \begin{equation}
                \label{eq:serie1}
                S_M = \sum_{i=0}^M \sum_{j=0}^M p(i,j),\quad p(x,y) = \frac{1}{1+a(x^2{+}y^2)^k},
            \end{equation}
            where $a>0$ and $k > 0$ are constants.
            Since $p(x,y)$ is decreasing in both variables, $p(i,j) \geq \iint_{Q_{ij}} p(x,y) \, \dd x \, \dd y$, with $Q_{ij} = [i,i+1] \times [j,j+1]$.

            Now, if $R_M$ is the circular sector $x \geq 0$, $y \geq 0$, $x^2+y^2 \leq M^2$, then in polar coordinates we have
            \begin{equation}
                \iint_{R_M} p(x,y) \, \dd x \, \dd y = \frac{\pi}{2} \int_0^M \!\! \frac{r}{1 + a\, r^{2k}} \,\dd r.
            \end{equation}
            The previous integral can be expressed in terms of the hypergeometric function, but we only need to know that in the limit $M \to +\infty$ the integral is finite if and only if $k>1$; and when $k=1$ its value is
            \begin{equation}
                \frac{\pi}{4a} \ln(1+aM^2) \sim \frac{\pi}{2a} \ln M \quad (M \to +\infty).
            \end{equation}
            Using this result it is not difficult to conclude that the series~\cref{eq:serie1} is convergent if and only if $k>1$.

            To finish the analysis we will see that when $k=1$ the divergence of the series is logarithmic, as the integral.
            Application of the first part of \cref{lem:series} twice shows that the difference between $S_M$ and
            \begin{equation}
                I_M = \iint_{Q_M}\!\! p(x,y) \dd x \dd y,\quad Q_M = [0,M]\times[0,M],
            \end{equation}
            is bounded by a constant independent of~$M$.
            Second, this integral can be expressed in polar coordinates and gives
            {\small\begin{equation}
                I_M = \frac{1}{2a}\bigg(\int_{0}^{\pi/4} \hspace{-2.25ex} \dd\phi \, \ln\Big(1 + \frac{aM^2}{\cos^2\phi}\Big)+\!\int_{\pi/4}^{\pi/2} \hspace{-2.25ex} \dd\phi \, \ln\Big(1 + \frac{aM^2}{\sin^2\phi}\Big)\!\!\bigg),
            \end{equation}}
            yielding $I_M \sim \frac{\pi}{2a} \ln M$ as $M \to +\infty$.
            The conclusion is stated as a lemma:
            \begin{lemma}
                \label{lem:double_series_1}%
                The series $\lim_{M \to +\infty} S_M$ given by~\cref{eq:serie1} is convergent if and only if $k>1$.
                For $k=1$, $S_M \sim \frac{\pi}{2a} \ln M$ as $M \to +\infty$.
            \end{lemma}

            A small variation of the preceding analysis yields:
            \begin{lemma}
                \label{lem:sum_no_const}%
                The sum
                \begin{equation}
                    \sum_{i,j=1}^M \frac{1}{(i^2{+}j^2)^k}
                \end{equation}
                is convergent when $M \to \infty$ iff $k>1$.
            \end{lemma}

            \medskip
            \noindent\textit{\ref{ap:lattice_sum}.2}
            The second sum of interest is
            \begin{equation}
                \label{eq:serie2}
                S_M = \sum_{i=1}^M \sum_{j=1}^M p(i,j),\quad p(x,y) = \frac{x^4}{(x^2+y^2)^3}.
            \end{equation}
            We will compare this sum with the integral
            \begin{equation}
                I_M = \iint_{Q_M}\!\! p(x,y) \dd x \dd y,\quad Q_M = [1,M]\times[1,M].
            \end{equation}
            The comparison is not so straightforward since $p(x,y)$ decreases as $y$ increases, but this does not hold for~$x$.
            For fixed~$y$, $p(x,y)$ increases for $x \leq \sqrt2 y$, then it decreases.
            So we will apply the two parts of \cref{lem:series}.
            We have:
            \begin{itemize}[itemsep=1ex]
                \item
                The difference between $\sum_{j=1}^M p(i,j)$ and the integral $\int_1^M \dd y \, p(i,y)$ is less than $p(i,1) \leq \frac{1}{i^2}$.
                \item
                Summing these differences for $1 \leq i \leq M$, the difference between $\sum_{i=1}^M \sum_{j=1}^M p(i,j)$ and $\int_1^M \dd y \, \sum_{i=1}^M p(i,y)$ is less than $\sum_{i=1}^M \frac{1}{i^2} \leq \pi^2/6$.
                \item
                The inner sum of the integral, $\sum_{i=1}^M p(i,y)$, differs from $\int_1^M \dd x \, p(x,y)$ less than $2\, p(\sqrt2 y,y) = \frac{8}{27 y^2}$.
                \item
                Integrating, we see that $\int_1^M \dd y \, \sum_{i=1}^M p(i,y)$ differs from $\int_1^M \dd y \int_1^M \dd x \, p(x,y)$ less than $\frac{8}{27}(1 - \frac 1M) < \frac{8}{27}$.
            \end{itemize}

            Summing up, $S_M$ differs from $I_M$ in less than $\frac{\pi^2}{6} + \frac{8}{27}$, so it only remains to compute $I_M$.
            Changing to polar coordinates one can show that
            \begin{equation}
                I_M = \ln M \int_{\arctan\frac 1M}^{\pi/4} \!\!\dd \phi \, (\cos^4\phi + \sin^4\phi) - c_M,
            \end{equation}
            where $0 < c_M < 1$.
            Taking into account that
            $\int_{0}^{\pi/4} \!\!\dd \phi \,(\cos^4\phi + \sin^4\phi) = \frac{3 \pi}{16}$,
            we have proved the following lemma:
            \begin{lemma}
                \label{lem:double_series_2}%
                The order of divergence of the series given by~\cref{eq:serie2} is $S_M \sim \frac{3 \pi}{16} \ln M$ as $M \to \infty$.
            \end{lemma}

        \subsection{Mutual Coupling Between Hertzian Dipoles}
            \label[appendix]{ap:mutual_z_hertzian}
            For a dipole of length $l$ along the $z$-axis, the open-circuit induced voltage can be computed as~\cite[Sec.~25.3]{orfanidis_electromagnetic_2016}:
            \begin{equation}
                \label{eq:v_21oc}
                V = -\int_{-l/2}^{l/2}E_z(z) \dd z.
            \end{equation}

            Now consider a single Hertzian dipole of length $l \ll \lambda$ located along the $z$-axis, with a uniform current $I$.
            Its electric field at a point (in spherical coordinates) $\mathbf{r}=(r, \theta, \phi)$ is given by~\cite[Sec.~4.2]{balanis_antenna_2016}:
            \begin{align}
                &E_r = \eta\frac{I\, l \cos\theta}{2\pi \kappa}\left(\frac{\kappa}{r^2}-\frac{\im}{r^3}\right)\euler^{-\im \kappa r},\\
                &E_{\theta} = \im\eta\frac{I\, l \sin\theta}{4\pi \kappa}\left(\frac{\kappa^2}{r}-\frac{\im \kappa}{r^2}-\frac{1}{r^3}\right)\euler^{-\im \kappa r},\\
                &E_{\phi} = 0.
            \end{align}
            If we add a second (open-circuited) dipole, the interaction between them depends on their relative orientation.
            Next, we derive the expression for the mutual impedance between two equally oriented Hertzian dipoles at arbitrary positions.

            Assume that the second dipole is placed at coordinates $(b, 0, h)$, so they are at a distance $r=\sqrt{b^2+h^2}$.
            Since a Hertzian dipole is electrically small ($l \ll \lambda$),
            \begin{equation}
                V \approx -E_z \cdot l = l\cdot(E_{\theta}\sin\theta - E_r\cos\theta).
            \end{equation}
            The mutual impedance is then given by
            \begin{equation}
                \label{eq:z_hertzian_general}
                Z(r,\theta) = \frac{V}{I} = \frac{3}{2}\R{r}\,\im\,\euler^{-\im\kappa r} \psi(r,\theta),
            \end{equation}
            where $\R{r} = \frac{2\pi}{3}\eta\left(\frac{l}{\lambda}\right)^2$ is the radiation resistance, and
            \begin{equation}
                \label{eq:psi_hertzian}
                \psi(r,\theta) = \frac{1-\cos^2\theta}{\kappa r} - \im\frac{1-3\cos^2\theta}{(\kappa r)^2}-\frac{1-3\cos^2\theta}{(\kappa r)^3}
            \end{equation}
            is an auxiliary function that captures propagation losses.
            Equivalently, this function can be expressed in terms of the Green function $g(r) = \frac{\euler^{-\im \kappa r}}{4\pi r}$ as
            \begin{equation}
                \label{eq:psi_green}
                \psi(r, \theta) = \frac{4\pi}{\kappa} \euler^{\,\im\kappa r} \left( 1 + \frac{1}{\kappa^2} \frac{\partial^2}{\partial z^2} \right) g(r).
            \end{equation}

            The above expression is valid for side-by-side, collinear or parallel-in-echelon dipoles.
            In the first case, $\theta=\pi/2$, and thus
            \begin{equation}
                Z_{||}(r) = \frac{3}{2}\R{r}\,\im\,\euler^{-\im \kappa r}\left(\frac{1}{\kappa r}-\frac{\im}{(\kappa r)^2}-\frac{1}{(\kappa r)^3}\right).
            \end{equation}
            Conversely, when the dipoles are collinear, $\theta=0$, and
            \begin{equation}
                Z_{|}(r) = 3\R{r}\,\euler^{-\im \kappa r}\left(-\frac{1}{(\kappa r)^2} + \frac{\im}{(\kappa r)^3}\right).
            \end{equation}

            Note that two side-by-side dipoles exhibit the strongest coupling, whereas two collinear dipoles exhibit the weakest.

        \subsection{Proof of Proposition~\ref*{prop:bound_D}}
            \label[appendix]{ap:proof_D}

            Let $\mathbf{G} = (\Z{G}\id{K}+\bfZ{in})^{-1}$ and $\mathbf{G}_\mathrm{UA} = (\Z{G}\id{K}+\bfZ{T})^{-1}$.
            We can write $\mathbf{G}$ in terms of $\mathbf{G}_\mathrm{UA}$ as
            \begin{equation}
                \begin{aligned}
                    &\mathbf{G} = (\Z{G}\id{K} + \bfZ{T} - \bfDelta{in})^{-1} = (\mathbf{G}_\mathrm{UA}^{-1} - \bfDelta{in})^{-1}\\
                    &= (\id{K}-\mathbf{G}_\mathrm{UA}\bfDelta{in})^{-1}\mathbf{G}_\mathrm{UA} = \sum_{k=0}^{\infty} (\mathbf{G}_\mathrm{UA}\bfDelta{in})^k \mathbf{G}_\mathrm{UA},
                \end{aligned}
            \end{equation}
            where the last equality follows from the Neumann series~\cite[Sec.~3.8]{meyer_matrix_2000}.
            Since $\mathbf{D}=\mathbf{Q}_\mathrm{UA}\bfZ{RT}\mathbf{G}$, we can write
            \begin{equation}
                \mathbf{D}-\mathbf{D}_\mathrm{UA} = \mathbf{D}_\mathrm{UA}\sum_{k=1}^{\infty}(\bfDelta{in}\mathbf{G}_\mathrm{UA})^k.
            \end{equation}
            A bound on the relative error is thus given by
            \begin{equation}
                \label{eq:D_error_bound}
                \frac{\norm{\mathbf{D}-\mathbf{D}_\mathrm{UA}}{2}}{\norm{\mathbf{D}_\mathrm{UA}}{2}} \leq \norm{\sum_{k=1}^{\infty}(\bfDelta{in}\mathbf{G}_\mathrm{UA})^k}{2} \leq \frac{\varepsilon_\mathrm{in}}{1-\varepsilon_\mathrm{in}},
            \end{equation}
            where the last step follows from the triangle inequality and the geometric series sum formula, assuming $\varepsilon_\mathrm{in} < 1$.

            Following a similar procedure with $\mathbf{Q}$ instead of $\mathbf{G}$, we obtain the second statement of the proposition in terms of $\varepsilon_\mathrm{out}$.

        \subsection{Proof of Proposition~\ref*{prop:bound_B}}
            \label[appendix]{ap:proof_B}

            Let $\mathbf{K} = (\id{K} - \mathbf{G}_\mathrm{UA}\bfDelta{in})^{-1}$.
            The power coupling matrix can be written as
            \begin{equation}
                \mathbf{B} = \R{G} \herm{\mathbf{G}_\mathrm{UA}} \herm{\mathbf{K}} (\Re[\bfZ{T}] - \Re[\bfDelta{in}])\mathbf{K}\mathbf{G}_\mathrm{UA}.
            \end{equation}
            Using the identity $\mathbf{K} = \id{K} + \mathbf{K}\mathbf{G}_\mathrm{UA}\bfDelta{in}$, we obtain
            \begin{equation}
                \mathbf{B} = \mathbf{B}_\mathrm{UA} + \R{G}\herm{\mathbf{G}_\mathrm{UA}}(\mathbf{T}_1 + \mathbf{T}_2 + \mathbf{T}_3 + \mathbf{T}_4)\mathbf{G}_\mathrm{UA},
            \end{equation}
            where $\mathbf{B}_\mathrm{UA} = \R{G}\herm{\mathbf{G}_\mathrm{UA}}\Re[\bfZ{T}]\mathbf{G}_\mathrm{UA}$ and
            \begin{equation}
                \begin{aligned}
                    &\mathbf{T}_1 = \herm{(\mathbf{K}\mathbf{G}_\mathrm{UA}\bfDelta{in})}\Re[\bfZ{T}],\\
                    &\mathbf{T}_2 = \Re[\bfZ{T}](\mathbf{K}\mathbf{G}_\mathrm{UA}\bfDelta{in}),\\
                    &\mathbf{T}_3 = \herm{(\mathbf{K}\mathbf{G}_\mathrm{UA}\bfDelta{in})}\Re[\bfZ{T}](\mathbf{K}\mathbf{G}_\mathrm{UA}\bfDelta{in}),\\
                    &\mathbf{T}_4 = -\herm{\mathbf{K}}\Re[\bfDelta{in}]\mathbf{K}.
                \end{aligned}
            \end{equation}

            We can now bound the relative error incurred by the unilateral approximation.
            Applying the triangle inequality yields
            \begin{equation}
                \label{eq:B_error}
                \frac{\norm{\mathbf{B}-\mathbf{B}_\mathrm{UA}}{2}}{\norm{\mathbf{B}_\mathrm{UA}}{2}} \leq \frac{1}{\norm{\mathbf{B}_\mathrm{UA}}{2}} \sum_{i=1}^{4}\norm{\R{G}\herm{\mathbf{G}_\mathrm{UA}}\mathbf{T}_i\mathbf{G}_\mathrm{UA}}{2}.
            \end{equation}
            The first term, corresponding to $i=1$, can be written as
            \begin{equation}
                \R{G}\herm{\mathbf{G}_\mathrm{UA}}\mathbf{T}_1\mathbf{G}_\mathrm{UA} = \herm{\mathbf{G}_\mathrm{UA}}\herm{(\mathbf{K}\mathbf{G}_\mathrm{UA}\bfDelta{in})}\mathbf{G}_\mathrm{UA}^{-\mathrm{H}}\mathbf{B}_\mathrm{UA}.
            \end{equation}
            Substituting this expression into~\cref{eq:B_error} and invoking the submultiplicability of the spectral norm results in
            \begin{equation}
                \label{eq:bound_T1}
                \begin{aligned}
                    &\frac{\norm{\R{G}\herm{\mathbf{G}_\mathrm{UA}}\mathbf{T}_1\mathbf{G}_\mathrm{UA}}{2}}{\norm{\mathbf{B}_\mathrm{UA}}{2}} \leq \norm{\mathbf{G}_\mathrm{UA}^{-1}\mathbf{K}\mathbf{G}_\mathrm{UA}\bfDelta{in}\mathbf{G}_\mathrm{UA}}{2}\\
                    &= \norm{(\id{K} - \bfDelta{in}\mathbf{G}_\mathrm{UA})^{-1}\bfDelta{in}\mathbf{G}_\mathrm{UA}}{2}\\
                    &\leq \norm{(\id{K} - \bfDelta{in}\mathbf{G}_\mathrm{UA})^{-1}}{2}\norm{\bfDelta{in}\mathbf{G}_\mathrm{UA}}{2} \leq \frac{\varepsilon_\mathrm{in}}{1-\varepsilon_\mathrm{in}}.
                \end{aligned}
            \end{equation}
            The last inequality follows from the Neumann series and the geometric sum, as in~\cref{eq:D_error_bound}.

            The second and third terms in~\cref{eq:B_error} exhibit a structure similar to the first one, so they can be bounded as
            \begin{align}
                \frac{\norm{\R{G}\herm{\mathbf{G}_\mathrm{UA}}\mathbf{T}_2\mathbf{G}_\mathrm{UA}}{2}}{\norm{\mathbf{B}_\mathrm{UA}}{2}} &\leq \frac{\varepsilon_\mathrm{in}}{1-\varepsilon_\mathrm{in}},\label{eq:bound_T2}\\
                \frac{\norm{\R{G}\herm{\mathbf{G}_\mathrm{UA}}\mathbf{T}_3\mathbf{G}_\mathrm{UA}}{2}}{\norm{\mathbf{B}_\mathrm{UA}}{2}} &\leq \frac{\varepsilon_\mathrm{in}^2}{(1-\varepsilon_\mathrm{in})^2}.\label{eq:bound_T3}
            \end{align}

            To bound the last term, we upperbound the numerator and lowerbound the denominator.
            In particular, applying submultiplicability along with $\norm{\Re[\bfDelta{in}]}{2} \leq \norm{\bfDelta{in}}{2}$ yields the numerator bound
            \begin{equation}
                \label{eq:numerator_bound}
                \norm{\R{G}\herm{\mathbf{G}_\mathrm{UA}}\mathbf{T}_4\mathbf{G}_\mathrm{UA}}{2} \leq \R{G} \norm{\mathbf{K}}{2}^2 \norm{\mathbf{G}_\mathrm{UA}}{2}^2 \norm{\bfDelta{in}}{2}.
            \end{equation}
            To lowerbound the denominator, since $\Re[\bfZ{T}]$ is Hermitian and positive definite, we let $\mathbf{v}$ denote the singular vector associated with the maximum singular value of $\mathbf{G}_\mathrm{UA}$, yielding
            \begin{equation}
                \label{eq:denominator_bound}
                \begin{aligned}
                    \norm{\mathbf{B}_\mathrm{UA}}{2} &\geq \herm{\mathbf{v}}\mathbf{B}_\mathrm{UA}\mathbf{v} = \R{G}\herm{\mathbf{v}}\herm{\mathbf{G}_\mathrm{UA}}\Re[\bfZ{T}]\mathbf{G}_\mathrm{UA}\mathbf{v}\\
                    &\geq \R{G}\lambda_\mathrm{min}(\Re[\bfZ{T}]) \norm{\mathbf{G}_\mathrm{UA}}{2}^2,
                \end{aligned}
            \end{equation}
            where the first inequality follows from the definition of the spectral norm, and the second from the bounds on the Rayleigh quotient~\cite[Sec.~7.5]{meyer_matrix_2000}.
            Dividing~\cref{eq:numerator_bound} by~\cref{eq:denominator_bound} results in
            \begin{equation}
                \label{eq:bound_T4}
                \begin{aligned}
                    &\frac{\norm{\R{G}\herm{\mathbf{G}_\mathrm{UA}}\mathbf{T}_4\mathbf{G}_\mathrm{UA}}{2}}{\norm{\mathbf{B}_\mathrm{UA}}{2}} \leq \norm{\mathbf{K}}{2}^2 \frac{\norm{\bfDelta{in}}{2}}{\lambda_\mathrm{min}(\Re[\bfZ{T}])}\\[0.75ex]
                    &\leq \frac{\norm{\bfDelta{in}}{2}}{(1-\varepsilon_\mathrm{in})^2\lambda_\mathrm{min}(\Re[\bfZ{T}])} = \frac{\varepsilon_\mathrm{in} \mu_\mathrm{T}}{(1-\varepsilon_\mathrm{in})^2},
                \end{aligned}
            \end{equation}
            with $\mu_\mathrm{T} = \sigma_\mathrm{min}(\bfZ{T})/\lambda_\mathrm{min}(\Re[\bfZ{T}]) \geq 1$.
            In~\cref{eq:bound_T4}, we used that $\sigmamin(\Z{G}\id{K}+\bfZ{T}) \geq \sigmamin(\bfZ{T})$ due to the passivity of the antenna array and the positive real part of $\Z{G}$.

            Finally, adding~\cref{eq:bound_T1,eq:bound_T2,eq:bound_T3,eq:bound_T4}, we obtain the unilateral approximation error bound
            \begin{equation}
                \frac{\norm{\mathbf{B}-\mathbf{B}_\mathrm{UA}}{2}}{\norm{\mathbf{B}_\mathrm{UA}}{2}} \leq \frac{(2+\mu_\mathrm{T}-\varepsilon_\mathrm{in})\varepsilon_\mathrm{in}}{(1-\varepsilon_\mathrm{in})^2},
            \end{equation}
            so the error on the power coupling matrix depends only on the input-side condition.

        \subsection{Proof of Proposition~\ref*{prop:bound_noise}}
            \label[appendix]{ap:proof_noise}

            Following a similar approach as with the power coupling matrix, we define an auxiliary matrix $\mathbf{J} = (\id{N}-\frac{\bfDelta{out}}{\Z{L}}\mathbf{Q}_\mathrm{UA})^{-1}$ satisfying $\mathbf{J} = \id{N}+\frac{\bfDelta{out}}{\Z{L}}\mathbf{Q}_\mathrm{UA}\mathbf{J}$ and $\mathbf{Q}=\mathbf{Q}_\mathrm{UA}\mathbf{J}$.
            We can now write noise covariance as
            \begin{equation}
                \cov{\bsf{n}} = \cov{\bsf{n},\mathrm{UA}} + \mathbf{Q}_\mathrm{UA}(\mathbf{R}_1+\mathbf{R}_2+\mathbf{R}_3+\mathbf{R}_4)\herm{\mathbf{Q}_\mathrm{UA}},
            \end{equation}
            where $\cov{\bsf{n},\mathrm{UA}} = \mathbf{Q}_\mathrm{UA}\mathbf{P}_\mathrm{UA}\herm{\mathbf{Q}_\mathrm{UA}}$ follows from~\cref{eq:noise_cov} setting $\bsfZ{out}=\bfZ{R}$ and
            \begin{equation}
                \begin{aligned}
                    &\mathbf{R}_1 = \Big(\frac{\bfDelta{out}}{\Z{L}}\mathbf{Q}_\mathrm{UA}\mathbf{J}\Big)\mathbf{P}_\mathrm{UA},\\
                    &\mathbf{R}_2 = \mathbf{P}_\mathrm{UA}\herm{\Big(\frac{\bfDelta{out}}{\Z{L}}\mathbf{Q}_\mathrm{UA}\mathbf{J}\Big)},\\
                    &\mathbf{R}_3 = \Big(\frac{\bfDelta{out}}{\Z{L}}\mathbf{Q}_\mathrm{UA}\mathbf{J}\Big)\mathbf{P}_\mathrm{UA}\herm{\Big(\frac{\bfDelta{out}}{\Z{L}}\mathbf{Q}_\mathrm{UA}\mathbf{J}\Big)},\\
                    &\mathbf{R}_4 = \mathbf{J}\boldsymbol{\Delta}\mathbf{P}\herm{\mathbf{J}},
                \end{aligned}
            \end{equation}
            with
            \begin{equation}
                \begin{aligned}
                    &\boldsymbol{\Delta}\mathbf{P} = \mathbf{P} \!-\! \mathbf{P}_\mathrm{UA} = \sigma_i^2(\bfDelta{out}\bfDelta[H]{out} + 2\R{N}\Re[\rho^*\bfDelta{out}]\\
                    &\qquad- \R{N}\Re[\bfDelta{out}] - \bfZ{R}\bfDelta[H]{out} - \bfDelta{out}\bfZ[H]{R}).
                \end{aligned}
            \end{equation}
            Observe how the $\mathbf{R}_i$ above are equivalent to the $\mathbf{T}_i$ in \cref{ap:proof_B}, so the approximation error will be
            \begin{equation}
                \frac{\norm{\cov{\bsf{n}}-\cov{\bsf{n},\mathrm{UA}}}{2}}{\big\Vert\cov{\bsf{n},\mathrm{UA}}\big\Vert_2} \leq \frac{(2+\mu_\mathrm{N}-\varepsilon_\mathrm{out})\varepsilon_\mathrm{out}}{(1-\varepsilon_\mathrm{out})^2},
            \end{equation}
            up to $O(\varepsilon _{\text{out}}^{2})$ in the numerator.

            To obtain $\mu_\mathrm{N}$ we focus on $\mathbf{R}_4$.
            Applying submultiplicability and the bounds on the Rayleigh quotient as in~\cref{eq:bound_T4}:
            \begin{equation}
                \hspace{-1ex}\begin{aligned}
                    &\frac{\norm{\mathbf{Q}_\mathrm{UA}\mathbf{R}_4\herm{\mathbf{Q}_\mathrm{UA}}}{2}}{\big\Vert\cov{\bsf{n},\mathrm{UA}}\big\Vert_2} \leq \frac{\norm{\boldsymbol{\Delta}\mathbf{P}}{2}}{(1-\varepsilon_\mathrm{out})^2 \lambda_\mathrm{min}(\mathbf{P}_\mathrm{UA})}\\
                    &\lesssim \frac{\varepsilon_\mathrm{out} \sigma_i^2 \sigmamin(\bfZ{R})(2\norm{\bfZ{R}}{2}+\R{N}(1+2|\rho|))}{(1-\varepsilon_\mathrm{out})^2 \lambda_\mathrm{min}(\mathbf{P}_\mathrm{UA})}\\
                    &\leq \frac{\varepsilon_\mathrm{out} \sigmamin(\bfZ{R})(2\norm{\bfZ{R}}{2}+\R{N}(1+2|\rho|))}{(1-\varepsilon_\mathrm{out})^2 \R{N}(\R{N}(1-|\rho|^2)+\lambda_\mathrm{min}(\Re[\bfZ{R}]))},
                \end{aligned}
            \end{equation}
            up to $O(\varepsilon _{\text{out}}^{2})$ in the numerator.
            Finally,
            \begin{equation}
                \label{eq:mu_N}
                \mu_\mathrm{N} = \frac{\sigmamin(\bfZ{R})(2\norm{\bfZ{R}}{2}+\R{N}(1+2|\rho|))}{\R{N}(\R{N}(1-|\rho|^2)+\lambda_\mathrm{min}(\Re[\bfZ{R}]))}.
            \end{equation}
            Note that $\mu_\mathrm{N}$ is an upper bound of the tighter expression $\sigma_i^2 \sigmamin(\bfZ{R})(2\norm{\bfZ{R}}{2}+\R{N}(1+2|\rho|))/\lambda_\mathrm{min}(\mathbf{P}_\mathrm{UA})$, obtained from a lower bound on $\lambda_\mathrm{min}(\mathbf{P}_\mathrm{UA})$.

        \subsection{Spatial Frequency Representation of Coupling Matrices}
            \label[appendix]{ap:spectral_representation}
            This appendix is devoted to deriving the spatial-frequency representations introduced in \cref{sec:los_multiuser_input} of the inter- and intra-array coupling matrices.

            Letting $(x_l, y_l, z_l)$ be the location of the $l$-th user, the continuum extension of the $l$-th inter-array coupling vector across the \acrshort{ap} aperture $(y,z)$ is given by
            \begin{equation}
                z_{\text{RT},l}(y,z) = \Gamma_l \frac{\euler^{-\im\kappa r}}{\kappa r} = \frac{\Gamma_l \euler^{-\im\kappa \sqrt{x_l^{2} + (y-y_l)^2 + (z-z_l)^2}}}{\kappa \sqrt{x_l^2 + (y\!-\!y_l)^2 + (z\!-\!z_l)^2}}.
            \end{equation}
            In order to compute its 2D Fourier transform, we apply the Weyl identity~\cite[Sec.~2.2]{chew_waves_1995} so that
            \begin{equation}
                \begin{aligned}
                    &z_{\text{RT},l}(y,z) = \Gamma_l \frac{-\im}{2\pi\kappa}\\
                    &\quad\times\iint_{\mathbb{R}^2} \!\!\!\!\dd k_y \dd k_z \frac{\euler^{\,\im\left(k_y(y-y_l)+k_z(z-z_l)-k_x(k_y,k_z)x_l\right)}}{k_x(k_y, k_z)}\\[1ex]
                    &\quad= \frac{-\im2\pi\Gamma_l}{\kappa} \mathcal{F}^{-1}_{\text{2D}}\left[\frac{\euler^{-\im k_x(k_y,k_z)x_l}\,\euler^{-\im(k_y y_l + k_z z_l)}}{k_x(k_y, k_z)}\right],
                \end{aligned}
            \end{equation}
            so it follows that
            \begin{equation}
                \label{eq:z_rt_spectral}
                \hspace{-0.75ex}\tilde{z}_{\text{RT},l}(k_y, k_z) = \frac{-\im2\pi\Gamma_l}{\kappa \cdot k_x(k_y, k_z)}\euler^{-\im (k_x(k_y,k_z)x_l + k_y y_l + k_z z_l)},
            \end{equation}
            where $k_x(k_y, k_z)$ is parametrized by the $y$ and $z$ wavenumber coordinates as
            \begin{equation}
                \label{eq:kx_wavenumber}
                k_x(k_y, k_z) =
                \begin{cases}
                    \sqrt{\kappa^2 - k_y^2 - k_z^2}, &k_y^2+k_z^2 \leq \kappa^2,\\
                    -\im\sqrt{k_y^2 + k_z^2 - \kappa^2}, &k_y^2+k_z^2 > \kappa^2.
                \end{cases}
            \end{equation}
            It is clear from~\cref{eq:z_rt_spectral} that $\tilde{z}_{\text{RT},l}(k_y, k_z)$ diverges as $1/k_x(k_y,k_z)$ on the boundary circle $k_y^2+k_z^2 = \kappa^2$.
            Although its square is not integrable in that region, this singularity is compensated in~\cref{eq:I_kl} by the corresponding divergence of $\lambda(k_y, k_z)$.

            Now, in order to compute $\mathbf{U} \mathbf{z}_{\text{RT},l}$ in \cref{eq:akl_1},\footnote{The vector $\mathbf{U}^* \mathbf{z}_{\text{RT},k}$ follows directly from reversing the sign of the spatial frequencies $(k_y,k_z) \to (-k_y,-k_z)$ in the conjugated Fourier matrix \(\mathbf{U}^{*}\).} we sample both domains such that
            \begin{equation}
                [\mathbf{U} \mathbf{z}_{\text{RT},l}]_{p,q} \!= \frac{1}{d^2\sqrt{N}}\smashoperator{\sum_{m,n\in\znumbers}}\! \tilde{z}_{\text{RT},l}\left(\!k_{y} + m \frac{2\pi}{d}, k_{z} + n\frac{2\pi}{d}\!\right).
            \end{equation}
            The visible region is constrained to the set
            \begin{equation}
                \label{eq:visible_region}
                \mathcal{V}(k_y, k_z) = \left\{ \left(k_y + m \frac{2\pi}{d}\right)^2 \!+ \left(k_z + n \frac{2\pi}{d}\right)^2\! \le \kappa^2 \right\}
            \end{equation}
            and given that the channel acts as a low-pass filter~\cite{pizzo_spatial_2022}, we can discard the higher-frequency replicas lying outside the visible region:
            \begin{equation}
                [\mathbf{U} \mathbf{z}_{\text{RT},l}]_{p,q} \approx \frac{1}{d^2\sqrt{N}} \tilde{z}_{\text{RT},l}(k_{y}, k_{z}),
            \end{equation}
            yielding the numerator and scaling factor in~\cref{eq:akl_approx}.

            The spectral representation of the intra-array coupling is obtained through a similar procedure.
            Extending $\bfZ{R}$ to continuous space\footnote{The continuous extension requires considering $\psi(r,\theta)$ without the discontinuity introduced by the loading coils, resulting in a singularity at $(y,z)=(0,0)$ corresponding to an infinite self-reactance.} we have
            \begin{equation}
                \label{eq:z_cont}
                Z(y, z) = \frac{3}{2} \R{r}\, \im\, \euler^{-\im\kappa r(y,z)}\, \psi(r(y,z), \theta(y,z)),
            \end{equation}
            with $r=\sqrt{y^2+z^2}$ and $\cos\theta = z/r$.

            Let $\hat{\lambda}(k_y, k_z)$ denote the eigenvalues of the impedance matrix obtained through the continuous extension (\ie without loading coils), with $k_y = p \cdot \frac{2\pi}{d\sqrt{N}}$ and $k_z = q\cdot \frac{2\pi}{d\sqrt{N}}$ ($p, q \in \mathbb{Z}$).
            Then, the Poisson summation formula~\cite[Sec.~7.3]{beerends_fourier_2003} yields
            \begin{equation}
                \label{eq:mu}
                \hat{\lambda}(k_y, k_z) = \frac{1}{d^2}\sum_{m,n \in \mathbb{Z}} \tilde{Z}\left(k_y + m \frac{2\pi}{d}, \, k_z + n \frac{2\pi}{d}\right).
            \end{equation}

            In~\cref{eq:mu}, $\tilde Z(k_y, k_z)$ is the 2D Fourier transform of $Z(y, z)$.
            It follows from the differentiation property of the Fourier transform, \cref{eq:psi_green,eq:z_cont} that
            \begin{equation}
                \tilde Z(k_y, k_z) = \left(\im \frac{6\pi R_r}{\kappa}\right)\left(\frac{\kappa^2-k_z^2}{\kappa^2}\right) \tilde{g}(k_y, k_z),
            \end{equation}
            with $\tilde{g}(k_y, k_z)$ the 2D Fourier transform of the Green function.
            The Weyl identity at $x=0$ yields
            \begin{equation}
                \frac{\euler^{-\im\kappa r}}{4\pi r} = \frac{-\im}{4\pi^2} \iint_{\mathbb{R}^2}\!\! \dd k_y \dd k_z\, \frac{\euler^{\,\im(k_y y + k_z z)}}{2k_x(k_y,k_z)},
            \end{equation}
            that corresponds to the inverse 2D Fourier transform of $\frac{-\im}{2k_x(k_y,k_z)}$.
            Hence,
            \begin{equation}
                \tilde Z(k_y, k_z) = \frac{3\pi R_r}{\kappa^3}\frac{\kappa^2-k_z^2}{k_x(k_y,k_z)}.
            \end{equation}

            Since $k_x(k_y,k_z)$ in~\cref{eq:kx_wavenumber} is real inside the visible region $\mathcal{V}(k_y, k_z)$ and purely imaginary outside it, $\tilde Z(k_y, k_z)$ is purely real on $\mathcal{V}(k_y, k_z)$ and purely imaginary elsewhere.

    \section*{Acknowledgment}
        \addcontentsline{toc}{section}{Acknowledgment}
        A. Martí thanks Núria Flores-Espinosa of KTH Royal Institute of Technology and Dr. David Raventós of Universitat Autònoma de Barcelona for the valuable discussions regarding the numerical simulations.

    \bibliography{IEEEabrv, references}

\begin{thebibliography}{10}
\providecommand{\url}[1]{#1}
\csname url@samestyle\endcsname
\providecommand{\newblock}{\relax}
\providecommand{\bibinfo}[2]{#2}
\providecommand{\BIBentrySTDinterwordspacing}{\spaceskip=0pt\relax}
\providecommand{\BIBentryALTinterwordstretchfactor}{4}
\providecommand{\BIBentryALTinterwordspacing}{\spaceskip=\fontdimen2\font plus
\BIBentryALTinterwordstretchfactor\fontdimen3\font minus
  \fontdimen4\font\relax}
\providecommand{\BIBforeignlanguage}[2]{{%
\expandafter\ifx\csname l@#1\endcsname\relax
\typeout{** WARNING: IEEEtran.bst: No hyphenation pattern has been}%
\typeout{** loaded for the language `#1'. Using the pattern for}%
\typeout{** the default language instead.}%
\else
\language=\csname l@#1\endcsname
\fi
#2}}
\providecommand{\BIBdecl}{\relax}
\BIBdecl

\bibitem{rappaport_wireless_2009}
T.~S. Rappaport, \emph{Wireless communications: principles and practice},
  2nd~ed. Upper Saddle River, NJ: Prentice Hall, 2009.

\bibitem{abadal_wave_2020}
S.~Abadal, C.~Han, and J.~M. Jornet, ``Wave propagation and channel modeling in
  chip-scale wireless communications: A survey from millimeter-wave to
  terahertz and optics,'' \emph{IEEE Access}, vol.~8, pp. 278--293, 2020.

\bibitem{hamza_wireless_2016}
A.~S. Hamza, J.~S. Deogun, and D.~R. Alexander, ``Wireless communication in
  data centers: A survey,'' \emph{IEEE Communications Surveys \& Tutorials},
  vol.~18, no.~3, pp. 1572--1595, 2016.

\bibitem{chafii_twelve_2023}
M.~Chafii, L.~Bariah, S.~Muhaidat, and M.~Debbah, ``Twelve scientific
  challenges for {6G}: Rethinking the foundations of communications theory,''
  \emph{IEEE Communications Surveys \& Tutorials}, vol.~25, no.~2, pp.
  868--904, 2023.

\bibitem{lu_tutorial_2024}
H.~Lu \emph{et~al.}, ``A tutorial on near-field {XL}-{MIMO} communications
  toward {6G},'' \emph{IEEE Communications Surveys \& Tutorials}, vol.~26,
  no.~4, pp. 2213--2257, 2024.

\bibitem{balanis_antenna_2016}
C.~A. Balanis, \emph{Antenna theory: Analysis and design}, 4th~ed. Hoboken,
  USA: Wiley, 2016.

\bibitem{mezghani_reincorporating_2024}
A.~Mezghani \emph{et~al.}, ``Reincorporating circuit theory into information
  theory,'' \emph{{IEEE} {BITS} the Information Theory Magazine}, vol.~4,
  no.~1, pp. 40--58, 2024.

\bibitem{zhu_electromagnetic_2024}
J.~Zhu, Z.~Wan, L.~Dai, M.~Debbah, and H.~V. Poor, ``Electromagnetic
  information theory: Fundamentals, modeling, applications, and open
  problems,'' \emph{IEEE Wireless Communications}, vol.~31, no.~3, pp.
  156--162, 2024.

\bibitem{ivrlac_toward_2010}
M.~T. Ivrlač and J.~A. Nossek, ``Toward a circuit theory of communication,''
  \emph{IEEE Transactions on Circuits and Systems I: Regular Papers}, vol.~57,
  no.~7, pp. 1663--1683, 2010.

\bibitem{laas_limits_2020}
T.~Laas, J.~A. Nossek, and W.~Xu, ``Limits of transmit and receive array gain
  in massive {MIMO},'' in \emph{2020 {IEEE} {Wireless} {Communications} and
  {Networking} {Conference} ({WCNC})}, 2020.

\bibitem{damico_holographic_2024}
A.~A. D'Amico and L.~Sanguinetti, ``Holographic {MIMO} communications: What is
  the benefit of closely spaced antennas?'' \emph{{IEEE} Transactions on
  Wireless Communications}, vol.~23, no.~10, pp. 13\,826--13\,840, 2024.

\bibitem{laas_reciprocity_2020}
T.~Laas, J.~A. Nossek, S.~Bazzi, and W.~Xu, ``On reciprocity in physically
  consistent {TDD} systems with coupled antennas,'' \emph{{IEEE} Transactions
  on Wireless Communications}, vol.~19, no.~10, pp. 6440--6453, 2020.

\bibitem{lu_compact_2026}
J.~Lu \emph{et~al.}, ``Compact ultra massive antenna arrays under mutual
  coupling: Modeling and spectral efficiency analysis,'' \emph{IEEE
  Transactions on Communications}, vol.~74, pp. 3174--3189, 2026.

\bibitem{marti_asymptotic_2024}
A.~Martí, J.~Riba, M.~Lamarca, and X.~Gràcia, ``Asymptotic analysis of
  near-field coupling in massive {MISO} and massive {SIMO} systems,''
  \emph{IEEE Communications Letters}, vol.~28, no.~8, pp. 1929--1933, 2024.

\bibitem{marti_unilateral_2026}
A.~Martí, L.~Sanguinetti, and X.~Gràcia, ``On the unilateral approximation
  condition with linear arrays in near-field {NLoS} propagation,'' in
  \emph{2026 29th International Workshop on Smart Antennas (WSA)}, 2026.

\bibitem{qian_physically_2026}
M.~Qian, L.~Ye, C.~Liu, H.~Wang, Y.~Song, and T.~Gao, ``Physically consistent
  near-field {MISO} communication: Impact of receiver-to-transmitter mutual
  coupling on achievable rate,'' \emph{IEEE Communications Letters}, vol.~30,
  pp. 2203--2207, 2026.

\bibitem{phang_near-field_2018}
S.~Phang, M.~T. Ivrlač, G.~Gradoni, S.~C. Creagh, G.~Tanner, and J.~A. Nossek,
  ``Near-field {MIMO} communication links,'' \emph{{IEEE} Transactions on
  Circuits and Systems I: Regular Papers}, vol.~65, no.~9, pp. 3027--3036,
  2018.

\bibitem{madhow_introduction_2014}
U.~Madhow, \emph{Introduction to communication systems}. Cambridge, UK:
  Cambridge University Press, 2014.

\bibitem{marti_physically_2026}
A.~Martí, ``Physically consistent wireless communications with statistical
  channel state information,'' Ph.D. dissertation, Universitat Politècnica de
  Catalunya, 2026.

\bibitem{meyer_matrix_2000}
C.~D. Meyer, \emph{Matrix analysis and applied linear algebra}, 1st~ed.
  Philadelphia, PA: Society for Industrial and Applied Mathematics, 2000.

\bibitem{pozar_microwave_2012}
D.~M. Pozar, \emph{Microwave engineering}, 4th~ed. Hoboken, {USA}: Wiley, 2012.

\bibitem{bensky_short-range_2019}
A.~Bensky, \emph{Short-range wireless communication}, 3rd~ed. Oxford, UK:
  Newnes, 2019.

\bibitem{bacci_spherical_2023}
G.~Bacci, L.~Sanguinetti, and E.~Björnson, ``Spherical wavefronts improve
  {MU}-{MIMO} spectral efficiency when using electrically large arrays,''
  \emph{IEEE Wireless Communications Letters}, vol.~12, no.~7, pp. 1219--1223,
  2023.

\bibitem{marti_harnessing_2025}
A.~Martí, L.~Sanguinetti, M.~Lamarca, and J.~Riba, ``Harnessing wavefront
  curvature and spatial correlation in noncoherent {MIMO} communications,''
  \emph{IEEE Wireless Communications Letters}, vol.~14, no.~8, pp. 2461--2465,
  2025.

\bibitem{marti_coherent_2025}
A.~Martí, L.~Sanguinetti, J.~Riba, and M.~Lamarca, ``Coherent and noncoherent
  detection in dense arrays: Can we ignore mutual coupling?'' in \emph{2025
  33rd European Signal Processing Conference (EUSIPCO)}, 2025, pp. 2027--2031.

\bibitem{wasylkiwskyj_theory_1970}
W.~Wasylkiwskyj and W.~K. Kahn, ``Theory of mutual coupling among
  minimum-scattering antennas,'' \emph{IEEE Transactions on Antennas and
  Propagation}, vol.~18, no.~2, pp. 204--216, 1970.

\bibitem{orfanidis_electromagnetic_2016}
S.~J. Orfanidis, \emph{Electromagnetic waves and antennas}, 2016.

\bibitem{gray_toeplitz_2005}
R.~M. Gray, ``Toeplitz and circulant matrices: A review,'' \emph{Foundations
  and Trends® in Communications and Information Theory}, vol.~2, no.~3, pp.
  155--239, 2005.

\bibitem{davis_circulant_1986}
P.~J. Davis, \emph{Circulant matrices}. New York, NY: Wiley, 1986.

\bibitem{bjornson_massive_2017}
E.~Björnson, J.~Hoydis, and L.~Sanguinetti, ``Massive {MIMO} networks:
  Spectral, energy, and hardware efficiency,'' \emph{Foundations and Trends®
  in Signal Processing}, vol.~11, no.~3, pp. 154--655, 2017.

\bibitem{lehmeyer_lna_2015}
B.~Lehmeyer, M.~T. Ivrlač, and J.~A. Nossek, ``{LNA} noise parameter
  measurement,'' in \emph{2015 {European} {Conference} on {Circuit} {Theory}
  and {Design} ({ECCTD})}, 2015.

\bibitem{bartle_introduction_2011}
R.~G. Bartle and D.~R. Sherbert, \emph{Introduction to real analysis}, 4th~ed.
  Hoboken, NJ: Wiley, 2011.

\bibitem{chew_waves_1995}
W.~C. Chew, \emph{Waves and fields in inhomogeneous media}. New York, NY: IEEE
  Press, 1995.

\bibitem{pizzo_spatial_2022}
A.~Pizzo, L.~Sanguinetti, and T.~L. Marzetta, ``Spatial characterization of
  electromagnetic random channels,'' \emph{IEEE Open Journal of the
  Communications Society}, vol.~3, pp. 847--866, 2022.

\bibitem{beerends_fourier_2003}
R.~J. Beerends, H.~G. {ter Morsche}, J.~C. {van den Berg}, and E.~M. {van de
  Vrie}, \emph{Fourier and {Laplace} transforms}. New York, NY: Cambridge
  University Press, 2003.

\end{thebibliography}
\end{document}